\documentclass[acmsmall, screen]{acmart}

\usepackage{color}
\usepackage{etoolbox}
\usepackage{listings}
\usepackage[T1]{fontenc}

\usepackage[scaled=0.8]{beramono}

\newcommand{\lstfontfamily}{\ttfamily}

\definecolor{darkviolet}{rgb}{0.5,0,0.4}
\definecolor{darkgreen}{rgb}{0,0.4,0.2} 
\definecolor{darkblue}{rgb}{0.1,0.1,0.9}
\definecolor{darkgrey}{rgb}{0.5,0.5,0.5}
\definecolor{lightblue}{rgb}{0.4,0.4,1}

\definecolor{stringColor}{rgb}{0.16,0.00,1.00}
\definecolor{annotationColor}{rgb}{0.39,0.39,0.39}
\definecolor{keywordColor}{rgb}{0.50,0.00,0.33}
\definecolor{commentColor}{rgb}{0.25,0.50,0.37}
\definecolor{javadocColor}{rgb}{0.25,0.37,0.75}
\definecolor{jTagColor}{rgb}{0.50,0.62,0.75}
\definecolor{eTagColor}{rgb}{0.50,0.62,0.75}
\definecolor{lineNumberColor}{rgb}{0.47,0.47,0.47}

\def\jTags{@author, @deprecated, @exception, @param, @return, @see, @serial, @serialData, @serialField, @since, @throws, @version}

\def\jAnnotations{
    classoffset=1,
    morekeywords={@Override, @Deperecated, @SuppressWarnings, @Retention, @Documented, @Target, @Inherited},
    keywordstyle=\color{annotationColor},
    classoffset=0
}

\def\eTags{FIXME, TODO, XXX}

\newrobustcmd{\markupJavadocs}[1]{%
\edef\mytok{\the\lst@token}%
\renewcommand*{\do}[1]{%
\ifdefstring{\mytok}{##1}%
{\color{jTagColor}\bfseries\listbreak}%
{}%
}%
{\color{javadocColor}%
\expandafter\docsvlist\expandafter{\jTags}%
\renewcommand*{\do}[1]{%
\ifdefstring{\mytok}{##1}%
{\color{eTagColor}\bfseries\listbreak}%
{}%
}%
\expandafter\docsvlist\expandafter{\eTags}%
#1}%
}%

\newrobustcmd{\markupComments}[1]{%
\edef\mytok{\the\lst@token}%
\renewcommand*{\do}[1]{%
\ifdefstring{\mytok}{##1}%
{\color{eTagColor}\bfseries\listbreak}%
{}%
}%
{\color{commentColor}%
\expandafter\docsvlist\expandafter{\eTags}#1}%
}%

\lstdefinestyle{eclipse}{
  basicstyle={\lstfontfamily},
  emphstyle=\bfseries,
  keywordstyle=\color{keywordColor}\bfseries,
  commentstyle=\markupComments,
  stringstyle=\color{stringColor},
  numberstyle=\color{lineNumberColor}\lstfontfamily,
  morecomment=[s][\markupJavadocs]{/**}{*/}, 
  showstringspaces=false,
  numbers=left,
}

\lstdefinestyle{black}{
  basicstyle=\small\lstfontfamily,
  numbers=left,
  columns=fullflexible,
  breaklines=true,
  mathescape=true,
  escapechar=\#,
  tabsize=4,
  frame=lines,
  showstringspaces=false
}

\lstdefinestyle{seminar}{
  basicstyle=\small\ttfamily,
  numbers=left,
  breaklines=true,
  mathescape=true,
  escapechar=\#,
  tabsize=4,
  showstringspaces=false
}

\expandafter\lstset\expandafter{\jAnnotations}
\usepackage{algorithm2e}
\usepackage{xcolor}

\SetAlFnt{\footnotesize}

\SetAlCapSty{xAlCapSty}

\SetCommentSty{xCommentSty}

\SetNlSty{mynlfont}{}{} 

\LinesNumbered

\SetSideCommentRight

\DontPrintSemicolon

\RestyleAlgo{algoruled}

\usepackage[autolanguage]{numprint}

\newcommand*\np[2][z]{
\ifx z#1%
$\numprint{#2}$%
\else%
$\numprint[#1]{#2}$%
\fi\xspace%
}

\usepackage{fp}
\newcommand{\ShowAbsoluteNumber}[1]{%
\ifnum #1<10%
{\hspace*{0pt}#1}%
\else%
\ifnum #1<100%
{\hspace*{0pt}#1}%
\else%
\ifnum #1<1000%
{\hspace*{0pt}#1}%
\else%
{\numprint{#1}}%
\fi%
\fi%
\fi%
}

\newcommand{\ShowPercentage}[2]{%
\FPeval\percentage{round(#1/#2*100,0)}%
\FPeval\percentageOneDecimal{round(#1/#2*100,1)}%
\ifnum \percentage=0%
{\np[\%]{\FPprint{percentageOneDecimal}}}%
\else%
\ifnum \percentage<10%
{\np[\%]{\FPprint{percentageOneDecimal}}}%
\else%
{\np[\%]{\FPprint{percentageOneDecimal}}}%
\fi%
\fi%
\xspace
}

\newcommand{\ShowPercentageTwo}[2]{%
\FPeval\percentagetwo{round(#1/#2*100,0)}%
\FPeval\percentageTwoDecimal{round(#1/#2*100,2)}%
\ifnum \percentagetwo=0%
{\np[\%]{\FPprint{percentageTwoDecimal}}}%
\else%
\ifnum \percentagetwo<10%
{\np[\%]{\FPprint{percentageTwoDecimal}}}%
\else%
{\np[\%]{\FPprint{percentageTwoDecimal}}}%
\fi%
\fi%
\xspace
}

\newlength\BARSIZE  
\newcommand{\inlinechart}[2]{%
\FPeval{\BLACKBARSIZE}{#1/#2}\textcolor{black!80}{\rule{\BLACKBARSIZE\BARSIZE}{1.6ex}}%
\FPeval{\BLACKBARSIZE}{1 - (#1/#2)}\textcolor{black!10}{\rule{\BLACKBARSIZE\BARSIZE}{1.6ex}}%
}

\newcommand*\percent[3][v]{%
\ifx q#1%
    \np{#2}/\np{#3}(\ShowPercentage{#2}{#3})\else%
\ifx p#1%
    \np{#2}(\ShowPercentage{#2}{#3})\else%
\ifx m#1%
    \np{#2}%
    \inlinechart{#2}{#3}\else%
\ifx t#1%
    \ShowPercentage{#2}{#3}%
    \inlinechart{#2}{#3}\else%
\ifx c#1%
    \inlinechart{#2}{#3}%
\else%
    \np{#2}%
    \ifx r#1%
        /\np{#3}%
    \fi%
    \hspace*{0.5ex}(\ShowPercentage{#2}{#3}) %
    \inlinechart{#2}{#3}%
    \xspace
\fi\fi\fi\fi\fi%
}

\usepackage{hyperref}
\usepackage{color}
\usepackage{tcolorbox}
\usepackage{balance}
\usepackage{makecell}
\usepackage{verbatim}
\usepackage{ulem}
\usepackage{mdframed}
\usepackage{cancel}
\usepackage{lipsum}
\usepackage{amsmath}
\usepackage{amssymb}
\usepackage{amsfonts}
\usepackage[font=small]{caption}
\usepackage{ifpdf}
\usepackage{graphicx}
\usepackage{pifont}

\usepackage[inkscapelatex=false]{svg}

\usepackage{algorithmic}
\usepackage{array}

\usepackage{caption}
\usepackage{multirow}
\usepackage{float}
\usepackage{mwe}
\newcommand{\code}[1]{\texttt{#1}}

\usepackage{url}
\usepackage{todonotes}
\usepackage{subfigure}
\usepackage{tikz}
\usepackage{xcolor}

\usepackage{booktabs,caption,fixltx2e}
\usepackage[flushleft]{threeparttable}

 \usepackage{enumitem}

\hypersetup{
 colorlinks=true,
 linkcolor=blue,
 citecolor=blue,
 filecolor=black,
 urlcolor=black,
 pdfauthor = {},
 pdftitle = {},
 pdfkeywords = {}
}

\definecolor{bg}{HTML}{F8F9FB}

\newcommand{\blackding}[1]{\ding{\numexpr181+#1\relax}}
\newcommand{\whiteding}[1]{\ding{\numexpr171+#1\relax}}

\AtBeginDocument{%
  }

\setcopyright{acmlicensed}
\copyrightyear{2018}
\acmYear{2018}
\acmDOI{XXXXXXX.XXXXXXX}

\acmJournal{JACM}
\acmVolume{37}
\acmNumber{4}
\acmArticle{111}
\acmMonth{8}

\begin{document}

\title[LLM-CompDroid]{LLM-CompDroid: Repairing Configuration Compatibility Bugs in Android Apps with Pre-trained Large Language Models}


\author{Zhijie Liu}
\email{liuzhj2022@shanghaitech.edu.cn}
\affiliation{%
  \institution{ShanghaiTech University}
  \city{Shanghai}
  \country{China}
}

\author{Yutian Tang}
\authornote{Yutian Tang (yutian.tang@glasgow.ac.uk) is the corresponding author.}
\email{yutian.tang@glasgow.ac.uk}
\affiliation{%
  \institution{University of Glasgow}
  \country{United Kingdom}
}

\author{Meiyun Li}
\authornote{Both authors contributed equally to this research.}
\email{meiyunli@chd.edu.cn}
\author{Xin Jin}
\authornotemark[2]
\email{jinxin@chd.edu.cn}
\affiliation{%
  \institution{Chang’an University}
  \city{Xi’an}
  \state{Shaanxi}
  \country{China}
}

\author{Yunfei Long}
\email{yl20051@essex.ac.uk}
\affiliation{%
  \institution{University of Essex}
  \country{United Kingdom}
}

\author{Liang Feng Zhang}
\email{zhanglf@shanghaitech.edu.cn}
\affiliation{%
  \institution{ShanghaiTech University}
  \city{Shanghai}
  \country{China}
}

\author{Xiapu Luo}
\email{csxluo@comp.polyu.edu.hk}
\affiliation{%
  \institution{The Hong Kong Polytechnic University}
  \city{Hong Kong SAR}
  \country{China}
}



\begin{abstract}
    XML configurations are integral to the Android development framework, particularly in the realm of UI display. However, these configurations can introduce compatibility issues (bugs), resulting in divergent visual outcomes and system crashes across various Android API versions (levels). In this study, we systematically investigate LLM-based approaches for detecting and repairing configuration compatibility bugs. Our findings highlight certain limitations of LLMs in effectively identifying and resolving these bugs, while also revealing their potential in addressing complex, hard-to-repair issues that traditional tools struggle with. Leveraging these insights, we introduce the LLM-CompDroid framework, which combines the strengths of LLMs and traditional tools for bug resolution. Our experimental results demonstrate a significant enhancement in bug resolution performance by LLM-CompDroid, with LLM-CompDroid-GPT-3.5 and LLM-CompDroid-GPT-4 surpassing the state-of-the-art tool, ConfFix, by at least 9.8\% and 10.4\% in both \textit{Correct} and \textit{Correct@k} metrics, respectively. This innovative approach holds promise for advancing the reliability and robustness of Android applications, making a valuable contribution to the field of software development.
\end{abstract}




\maketitle

\section{Introduction}\label{sec:Introduction}

XML configurations are crucial to the platform's development framework in both Android apps and the Android ecosystem. They offer a structured, human-readable format for specifying various parameters essential for the accurate rendering of Android apps. These configurations encompass a broad spectrum of elements. Each element is designed to delineate a particular aspect of the app, such as user interface layouts, data resources, required permissions, and manifest declarations. 

\begin{figure}[h]
    \centering
    \subfigure{
    \includegraphics[width=0.32\textwidth]{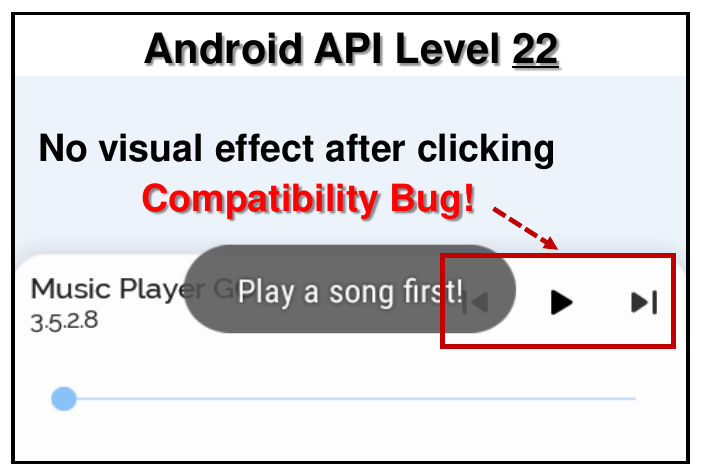}}
    \subfigure{
    \includegraphics[width=0.32\textwidth]{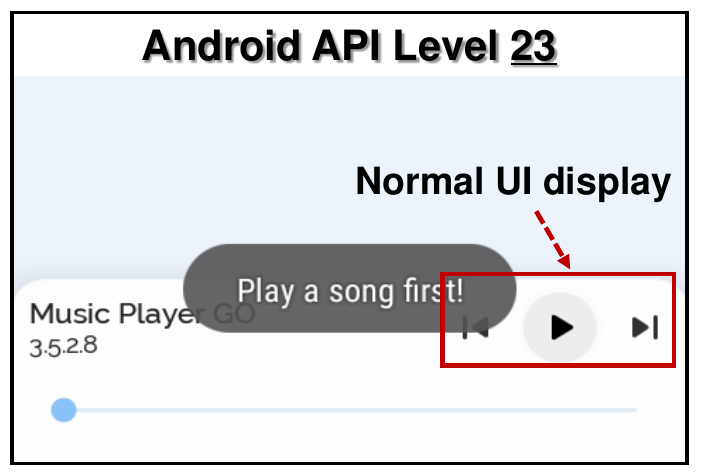}}
    \subfigure{
    \includegraphics[width=0.32\textwidth]{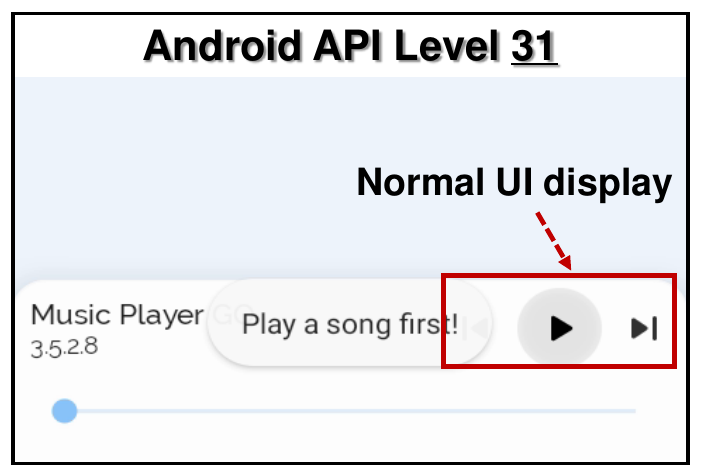}}
    \caption{Example of compatibility bug across Android API levels 22, 23, and 31.}
    \label{fig:example-compatibility-bug-intro}
\end{figure}

Despite their foundational role, XML configurations can also introduce configuration compatibility bugs (i.e., configuration compatibility issues) due to the fragmentation of the Android ecosystem. XML configurations can lead to configuration compatibility bugs when applications encounter discrepancies in resource availability, layout rendering, and behavior interpretation across diverse Android versions and device dimensions. These disparities arise from variations in XML schema interpretations, deprecated attributes, and evolving design guidelines. For example, Fig. \ref{fig:example-compatibility-bug-intro} illustrates a compatibility bug in an app named Music-Player-GO~\cite{Music:aef85dc}. Music-Player-GO is a free and open-source music player for Android 5.0 and above. As illustrated in Fig. \ref{fig:example-compatibility-bug-intro}, the visual effects after clicking the play button are different.

\noindent\textbf{State-of-the-art and Motivation.} To cope with compatibility bugs (issues)\footnote{There are two types of compatibility bugs: Android API compatibility bugs and configuration compatibility bugs. The former is introduced by the Android API inconsistency at the code level across different Android API levels.}~\cite{He:2018, Li:2018, Li:2018cid, Liu:2022, Wei:2016, Wei:2018, Wei:2019} including configuration compatibility bugs~\cite{Huang:2021,Huang:2023} which we focus on, researchers propose various approaches to detect compatibility bugs and repair those bugs in Android apps. To detect compatibility bugs in Android apps, researchers propose dynamic-based approaches (i.e., utilizing test cases \cite{Fazzini:2017,Ki:2019}) and static-based approaches (i.e, pre-defined rules \cite{Huang:2018,Huang:2021,Huang:2023,He:2018,Li:2018,Li:2018cid,Liu:2022,Wei:2016,Wei:2018,Wei:2019}) to detect potential compatibility bugs in apps. In terms of repairing these compatibility bugs, researchers propose a series of techniques \cite{Huang:2023,Zhao:2022,Lamothe:2020,Haryono:2020,Haryono:2022androevolve,Fazzini:2019} to cope with compatibility bugs, such as template-based approach, extracting knowledge from Android systems. In summary, for dynamic-based approaches, if certain paths or related information are not explored, compatibility bugs specific to that information might remain undetected. For static-based approaches, they lack the ability to understand the full context and semantics of the code. It might miss certain compatibility bugs that require understanding the runtime behavior and interactions with other components. However, using large language models can complement these approaches by providing a more human-friendly and contextually rich understanding of potential compatibility problems, and in this paper, particularly, we focus on the realm of configuration compatibility bugs.

\noindent\textbf{Pre-trained Large Language Models (LLMs).} In recent years, the realm of artificial intelligence has witnessed remarkable advancements, particularly in the field of natural language processing. One prominent development in this arena is the emergence of large language models (LLMs). These models represent a groundbreaking approach to comprehending and generating human language, enabled by their massive neural architectures and extensive training on vast corpora of text data. LLMs exhibit the capacity to grasp intricate syntactic and semantic nuances within language, allowing them to perform an array of tasks such as text generation, translation, summarization, and even engaging in contextual conversations. Their ability to understand context, context shifts, and incorporate diverse writing styles signifies a substantial leap towards machines emulating human-like linguistic prowess. As the software engineering landscape continues to evolve, the integration of LLMs holds the prospect of transforming the way software is conceived, crafted, and managed, marking a significant leap toward more efficient and effective software development practices. LLMs play multifaceted roles in revolutionizing various dimensions of software engineering tasks.

\noindent\textbf{Our Solution.} To cope with the configuration compatibility bugs in Android apps, in this paper, we first conduct a systematical study to evaluate the performance of 3 existing representative LLMs (i.e., GPT-3.5, GPT-4, and Bard) on detecting and repairing compatibility bugs in Android apps. Then, we propose a novel framework named LLM-CompDroid, which takes the advantages from LLMs and dynamic analysis (traditional tools) to detect and repair these bugs in apps.

\noindent\textbf{Contribution.} In summary, we make the following contributions in this paper:

\noindent$\bullet$ To the best of our knowledge, this is the \emph{first} paper conducting a systematical study on evaluating the performances of 3 existing LLMs on detecting and repairing configuration compatibility bugs;

\noindent$\bullet$ In this paper, we propose a novel framework called \textbf{LLM-CompDroid} to repair configuration compatibility bugs by combining the advantages from the LLMs and dynamic analysis; and

\noindent$\bullet$ The experiments on configuration compatibility bug repairing show that our \textbf{LLM-CompDroid} outperform the existing state-of-the-art solution (i.e., ConfFix~\cite{Huang:2023}) by at least 9.8\% and 10.4\% in \textit{Correct} and \textit{Correct@k} metrics, respectively.

\noindent \textbf{Online Artifact.} The data and experimental results are available at: \cite{artifact}.
\section{Background and Preliminary}\label{sec:Background}
\subsection{Configuration Compatibility Bugs in Android Apps}\label{sec:Configuration Compatibility Bug}
Compatibility bugs, also known as compatibility issues or compatibility-related defects, refer to software glitches that arise when a program, application, or system fails to function correctly or as intended due to inconsistencies or conflicts between different components, platforms, or environments \cite{Scalabrino:2019,Khoshnoud:2022,Zhao:2022,Huang:2023,Huang:2021}. These bugs typically occur when software elements interact in unexpected ways, leading to errors, crashes, or malfunctions. Compatibility bugs can manifest in various forms, including graphical anomalies, incorrect behavior, or performance degradation.

Compatibility bugs in Android apps can result in app crashes, distorted user interfaces, performance slowdowns, and unexpected behavior. In the context of Android apps, compatibility bugs specifically refer to two main types of compatibility bugs: configuration compatibility bugs \cite{Huang:2023, Huang:2021} and Android API compatibility bugs~\cite{Zhao:2022, Huang:2018}. The former is related to the XML elements and attributes in XML configuration files (\code{/res/} directory) and the latter is related to the Android API inconsistency at the code level across different Android API levels.

Specifically, a configuration compatibility bug $b$ is defined as a triplet <\textsf{issue-inducing XML element} $e_b$, \textsf{issue-inducing attributes} $A_b$, \textsf{conflicting Android API levels} $C_b$>. $A_b$ is a set of issue-inducing attributes located in $e_b$, an issue-inducing XML element. Additionally, the conflicting Android API levels $C_b$ is defined as a tuple <$l$, $l+1$> where inconsistent visual effect (UI rendering) or crashing is induced across Android API levels $l$ and $l + 1$ due to $A_b$ and $e_b$. A $b$ occurs when the Android framework parses $e_b$, loads $A_b$ to an UI object, or uses $A_b$'s values to render the UI object's runtime behavior across two neighboring $C_b$~\cite{Huang:2023} (an example of configuration compatibility bug is shown in Sec. \ref{sec:motivationexample}). This definition of the configuration compatibility bug follows the same one in \cite{Huang:2021} and \cite{Huang:2023}.

\subsection{Pre-trained Large Language Models (LLMs)}
Pre-trained large language models (LLMs) exhibit the capacity to grasp intricate syntactic and semantic nuances within language, allowing them to perform an array of tasks such as text generation, summarization, and even engaging in contextual conversations~\cite{Carlini:21, Brants:07, Raffel:22, nagata2021exploring}. Their ability signifies a substantial leap towards machines emulating human-like linguistic prowess. Beyond their accomplishments in natural language understanding and generation, LLMs have exhibited remarkable potential in addressing intricate challenges within the domain of software engineering~\cite{liu2023fill, pearce:2023, liu:2023-no, tang:2023, lemieux2023codamosa, deng2023large, fan2023automated, xia2023universal}. Leveraging their capacity to comprehend and analyze complex technical documentation, codebases, and software-related discussions, LLMs offer a promising avenue for automating various software engineering tasks. From bug detection and code completion to generating documentation and aiding in software maintenance, LLMs have showcased their ability to comprehend the nuances of programming languages and coding practices. Their capability to glean insights from a plethora of code examples, programming forums, and software repositories enables them to provide intelligent solutions.

Exploring the potential of utilizing LLMs for detecting and repairing configuration compatibility bugs in Android apps could be a promising avenue for investigation. The reasons can be explained as follows: (1) LLMs can learn from a huge amount of data. As they encounter more and more examples of these bugs and their fixes, they might become even better at identifying configuration compatibility bugs; and (2) the traditional tools that use certain rules to find configuration compatibility bugs may not identify configuration compatibility bugs that do not conform to pre-defined rules, which is also the same for repairing. However, LLMs can potentially perform better when dealing with such previously unseen data. Therefore, it is worth investigating how LLMs perform in detecting and repairing configuration compatibility bugs in Android apps.

\subsection{Motivating Example}\label{sec:motivationexample}
We leverage a running example (Fig. \ref{fig:example-compatibility-bug-intro}) to illustrate how an LLM (GPT-4) can be used to detect and repair a compatibility bug in an app. Refer to Fig. \ref{fig:example-compatibility-bug-intro}, the bug-related $e_b$ is <\code{ImageView}> that contains an $A_b$ of \code{android:foreground}. However, $A_b$ is introduced for $e_b$ in Android API level 23, which means that $b$ occurs across $C_b$ of <22, 23>~\cite{Android-Developers}, resulting in different visual effects across different Android API levels. We first ask GPT-4 whether there is any configuration compatibility bug in $e_b$, and it accurately identifies $b$ and the cause of $b$. Then, we require GPT-4 to repair $b$ and GPT-4 responds with a correct patch for repairing $b$ by introducing <\code{FrameLayout}> element to wrap original $e_b$ and move $A_b$ into the new element (see Fig. \ref{fig:repaired-music}). GPT-4 successfully repairs the bug which cannot be repaired by traditional tools~\cite{Huang:2023, Lint, XFix}.

\begin{figure}[h]
    \centering
  \begin{lstlisting}[language=XML]
+   <FrameLayout ...
+       android:foreground="?android:attr/actionBarItemBackground">
        <ImageView ...
-           android:foreground="?android:attr/actionBarItemBackground" 
        .../>
+   </FrameLayout>
  \end{lstlisting}

  \caption{Repared result by GPT-4 for Music-Player-GO.}
  \label{fig:repaired-music}
\end{figure}


\section{A Systematic Study on LLMs for Detecting and Repairing Configuration Compatibility Bugs}\label{sec:EmpricialStudy}

Prior to defining our own approach to detect and repair configuration compatibility bugs for Android apps, it is worth investigating how the existing LLMs perform directly on detecting and repairing these bugs in apps.

\subsection{Subject LLMs}
To evaluate the performance of existing pre-trained LLMs, we carefully select 3 representative LLMs: GPT-3.5 \cite{OpenAI}, GPT-4 \cite{GPT-4}, and Google Bard \cite{Bard}:


\noindent$\bullet$ \textbf{GPT-3.5}: The GPT-3.5 is the default LLM model offered by OpenAI and used in ChatGPT. It contains 175 billion parameters, making it a highly capable and complex model. GPT-3.5 is designed to perform a wide variety of natural language processing tasks, including text generation, text completion, and so forth. In this study, we utilize the model version \textsf{gpt-3.5-turbo}~\cite{GPT-Model}\footnote{At the time of experimenting, \textsf{gpt-3.5-turbo} corresponds to \textsf{gpt-3.5-turbo-0613}.} of GPT-3.5 for performing evaluation. We query it through OpenAI API~\cite{OpenAI-API}; 


\noindent$\bullet$ \textbf{GPT-4}: The GPT-4 is the state-of-the-art LLM model offered by OpenAI. The model contains over 1.76 trillion parameters. It is designed to perform a wide variety of natural language processing tasks similar to GPT-3.5. In addition, GPT-4 demonstrates much more powerful capabilities in certain complex scenarios~\cite{GPT-4, GPT-Model} than GPT-3.5. In this study, we utilize model version \textsf{gpt-4}~\cite{GPT-Model}\footnote{At the time of experimenting, \textsf{gpt-4} corresponds to \textsf{gpt-4-0613}.} as the representative of GPT-4. We query \textsf{gpt-4} through OpenAI API~\cite{OpenAI-API}; and

\noindent$\bullet$ \textbf{Google Bard}: Bard is a large language model (LLM) chatbot developed by Google AI. It was announced in February 2023, and is still under development. It is trained on a massive dataset of text and code, and can generate text, translate languages, write different kinds of creative content, and so forth. The model has 137 billion parameters. Google offers only one Bard model version which is accessible through web browsing~\cite{Bard}.

\subsection{Research Questions}
The systematical study on the performances of the existing LLMs is led by answering the following research questions (RQs):

\noindent$\bullet$ \textbf{RQ 1:} Can LLMs detect configuration compatibility bugs in Android Apps?

\noindent$\bullet$ \textbf{RQ 2:} Can LLMs identify conflicting Android API levels for compatibility bugs?  

\noindent$\bullet$ \textbf{RQ 3:} How well do LLMs perform at repairing configuration compatibility bugs?

The details of these RQs are discussed from Sec. \ref{subsec:rq1} to Sec. \ref{subsec:rq3}.

\subsection{Dataset and Study Setups}\label{sec:dataset-setup}
These research questions are conducted with the following dataset and experimental setups.

\noindent\textbf{Dataset:} In this study, to reduce potential bias in building our own dataset, we reuse the well-studied dataset presented in paper \cite{Huang:2023} for our study. The dataset contains 13 Android apps on GitHub~\cite{GitHub} with 77 compatibility bugs (the nominal ranges of apps’ versions are considered to these bugs) in total, which are detected by the state-of-the-art configuration compatibility bug detection tool ConfDroid~\cite{Huang:2021}. The 13 Android apps are diverse in multiple app categories and popular with thousands to millions of downloads in Google Play. Each of them is with rich maintenance and most of them have over 1,000 stars on GitHub. Moreover, the 77 compatibility bugs also cover diverse issue-inducing attributes in XML elements with compatibility bugs, ranging from Android API levels from 21 to 31. Specifically, there are 10 distinct ones. The definition of a compatibility bug can be found in Sec. \ref{sec:Configuration Compatibility Bug}.

\noindent\textbf{Experimental Setups:} To evaluate the performance of LLMs on detecting and repairing configuration compatibility bugs for Android apps, we adopt various metrics carefully which are appropriate for each research question's evaluation. These metrics are introduced specifically in each research question (see Sec. \ref{subsec:rq1} to Sec. \ref{subsec:rq3}). 

We set the applicable maximum Android API level to 31 (< the latest level 34~\cite{Android-API-Level}) due to that our selected GPT-3.5 and GPT-4 models have limited knowledge after 2021\footnote{Android API level 31 is released on October 4, 2021. Android API level 32 is released on March 7, 2022.}~\cite{GPT-Model} (there is no specific information for Bard). In addition, there is no compatibility bug with conflicting Android API levels in 32 to 34 in the dataset. Therefore, the maximum setting has no effect on the experimental results. We also set the applicable minimum one to 21 by following the same setting in \cite{Huang:2023}.

\noindent\textbf{Principle of Prompt Design:} The prompt design objective is not centered around discovering the perfect prompt that maximizes the performance of LLMs. Instead, our goal is to provide a reasonable prompt that can be used to evaluate the performance of LLMs in a general condition. By observing recent LLM-related research~\cite{liu2023fill, liu:2023-chatting, tang:2023, pearce:2022, pearce:2023, liu:2023-no, xia:2023}, we establish the following principle of prompt design: \textit{offer \emph{adequate but not excessive} information to LLMs for detecting and repairing configuration compatibility bugs}. The designed prompt template is defined as follows:

\begin{lstlisting}[
    basicstyle=\ttfamily\footnotesize,
    xleftmargin=0.5ex,
    backgroundcolor=\color{bg},
    breaklines=true,
    numbers=none,
    escapeinside=||
]
|\textbf{Prompt}:|
|Background:|
|<Introduction for Configuration Compatibility Bugs> \blackding{1}|
|<LLM> \blackding{2} Task:|
|<Task Description> \blackding{3}|
```xml
|<Issue-inducing XML Element> \blackding{4}|
```
|<LLM> \blackding{2} Response:|
|<Response Requirement> \blackding{5}|
\end{lstlisting}

\noindent Where <*> represents the information that needs to be filled in based on the configuration compatibility bugs. \blackding{1} is the basic configuration compatibility bug information. We provide this information to LLMs to enable them to comprehend the tasks they need to accomplish subsequently. This information includes an explanation of what configuration compatibility bugs are, how they arise, and the consequences they lead to. The specific text content can be found in our artifact. \blackding{2} represents the specific LLM used when prompting. \blackding{3} introduces the tasks that LLMs need to complete. There are a total of three tasks, corresponding to RQ1, RQ2, and RQ3, respectively. \blackding{4} is the issue-inducing XML element that is to be detected or fixed. \blackding{5} specifies the response requirements related to tasks, outlining the expected output content and format from the LLMs. In addition, the constructed prompts based on configuration compatibility bugs (in the dataset) and tasks are in a token count of less than 1,000, which does not hurt the output responses of LLMs\footnote{\textsf{gpt-3.5-turbo} and \textsf{gpt-4} have maximum tokens of 4,096 and 8,192, respectively. Bard's maximum token number is over 10,000.}.

\noindent\textbf{Environment:} All experiments are conducted on a server with an Intel(R) Core(TM) i9-10900X CPU @ 3.70GHz and 128GB RAM. The operating system is Ubuntu 20.04. The scripts used in the experiments are developed in Python 3.10. We follow the setting of Android emulators as \cite{Huang:2023} for evaluation, avoiding inconsistencies in the device (i.e., hardware) dimension. The Android emulators have 4 cores, 4GB RAM, 8GB internal storage, and a screen resolution of 1440 $\times$ 3120. 



\subsection{RQ 1: Can LLMs detect configuration compatibility bugs in Android Apps?}\label{subsec:rq1}

\noindent\textbf{Motivation.} In this RQ, we intend to evaluate whether LLMs can detect configuration compatibility bugs in Android apps. By answering this RQ, we are able to evaluate whether using LLMs to detect configuration comparability bugs in apps is a reasonable solution.

\noindent\textbf{Methodology.} We have LLMs directly detect the 77 configuration compatibility bugs to evaluate whether LLMs can discover these bugs in issue-inducing XML elements. This experiment only focuses on recognizing bugs in issue-inducing XML elements. We do not perform detection on all XML elements in Android apps since it is not scalability in time and cost dimension due to LLMs' query rate limits and token pricing~\cite{OpenAI}. Moreover, there is also no applicable way by using bug-related information (e.g., GitHub issue~\cite{GitHub-Issues}) to reduce the detection space of XML elements. 

We only utilize the issue-inducing XML elements to fill in the prompt and set the LLM task to whether the provided XML elements and their attributes can cause configuration compatibility bugs across Android API levels. We initially do not set the minimum Android API level to 21 in the prompt. We also specify that LLMs need to provide issue-inducing attributes (attribute context: AC) in these XML elements if the responses are \textsf{[Yes]} (XML element context: XEC).

We employ two metrics to assess the detection performance: Recall and Precision. Recall represents either the rates of truly detected compatibility bugs (i.e., true positives) or truly recognized issue-inducing attributes to the ground truth by LLMs, based on XEC or AC. Precision represents the rate of truly recognized issue-inducing attributes to all recognized attributes. We set both GPT-3.5 and GPT-4's temperatures to 0 to stabilize their responses. Thus, we only query them with 1 run for each prompt. However, Bard does not provide the option of temperature. We make it perform 1 run for each prompt like GPT-3.5 and GPT-4 to maintain consistency in the setting. 



\begin{table}[t]
\centering
\caption{Configuration Compatibility Detection Results by GPT-3.5, GPT-4, and Bard}

\scalebox{0.8}{\begin{tabular}{c|r|r|r|r} 
\toprule
\multicolumn{1}{c|}{\multirow{2}{*}{\textbf{LLM}}} & \multicolumn{1}{c|}{\textbf{XML Element}} & \multicolumn{3}{c}{\textbf{Attribute}}   \\ 

\cline{2-5}

& \multicolumn{1}{c|}{\textbf{Recall}} & \multicolumn{1}{c|}{\textbf{Recall}}  & \multicolumn{1}{c|}{\textbf{Precision}} & \multicolumn{1}{c}{\textbf{Avg. Precision}} \\

\hline

GPT-3.5 & \percent[r]{54}{77}  &  \percent[r]{49}{77}  &  \percent[r]{49}{251} &  \percent[t]{27.6}{100} \\

\hline

GPT-4 & \percent[r]{56}{77} & \percent[r]{46}{77} &  \percent[r]{46}{146} &  \percent[t]{41.9}{100} \\

\hline

Bard & \percent[r]{75}{77} & \percent[r]{36}{77} & \percent[r]{36}{93} & \percent[t]{45.0}{100} \\

\bottomrule
\end{tabular}}
\label{tab:ccdr}
\end{table}

\noindent\textbf{Results.} The compatibility detection results by three LLMs are shown in Table \ref{tab:ccdr}. GPT-3.5 and GPT-4 only achieve similar 70.1\% and 72.7\% Recall values in XEC, respectively. Bard can get the best compatibility detection performance of Recall value with 97.4\%, surpassing GPT-3.5 and GPT-4 by 27.3\% and 24.7\%, respectively. Nevertheless, when it comes to issue-inducing attribute detection, Bard is weaker compared to GPT-3.5 and GPT-4. Specifically, Bard only achieves 46.8\% Recall value, and GPT-3.5 and GPT-4 get 63.6\% and 59.7\% Recall values, respectively. The recall values for all three LLMs in AC are below 65\%, indicating suboptimal performance in detecting issue-inducing attributes. Even worse, all three LLMs introduce a significant number of false positive attributes. Their Precision values are all below 40\%, where the lowest one is GPT-3.5's 19.5\%. GPT-3.5 detects 251 issue-inducing attributes but only 49 of them are the real ones. GPT-4 detects 100 false positives. In comparison, Bard detects the lowest number of false positives, with only 57. We also calculate the Avg. (average) Precision of all configuration compatibility bugs'. The results are displayed in the fifth column of Table \ref{tab:ccdr}. GPT-3.5, GPT-4, and Bard achieve 27.6\%, 41.9\%, and 45.0\%, respectively. 

\begin{table}[t]
\centering
\caption{Issue-inducing Attributes' Mean and Std Dev of Corresponding Precision values of Configuration Compatibility Bugs (No. 3, 5, and 8 are canceled since they correspond to the bugs with only one occurrence)}
\scalebox{0.8}{\begin{tabular}{c|c|c|c|c|c|c|c|c|c|c|c} 
\toprule
\multirow{2}{*}{\textbf{LLM}} & \multirow{2}{*}{\textbf{Metric}} & \multicolumn{10}{c}{\textbf{Attribute}}   \\

\cline{3-12}
& & No. 1& No. 2& \cancel{No. 3} & No. 4& \cancel{No. 5} & No. 6& No. 7& \cancel{No. 8} & No. 9& No. 10 \\

\hline

\multirow{2}{*}{GPT-3.5} & Mean & 0.25& 0.67& \cancel{-} & 0.11 & \cancel{-} & 0.11 &  0.17 & \cancel{-} & 0.23& 0.25 \\

\cline{2-12}

& Std Dev & 0.24 & 0.24  & \cancel{-} & 0.01 & \cancel{-} & 0.18 & 0.0 & \cancel{-} & 0.02& 0.0 \\
\hline

\multirow{2}{*}{GPT-4} & Mean & 0.37 & 1.0& \cancel{-}& 0.0& \cancel{-} & 0.06& 0.70& \cancel{1.0}& 0.18& 0.42 \\

\cline{2-12}

& Std Dev & 0.23 & 0.0& \cancel{-} & 0.0& \cancel{0.0}& 0.12& 0.24& \cancel{0.0}& 0.07& 0.08 \\
\hline

\multirow{2}{*}{Bard} & Mean & 0.63& 1.0& \cancel{0.0}& 0.03& \cancel{0.0}& 0.33& 0.0& \cancel{1.0}& 0.14  & 1.0 \\

\cline{2-12}

& Std Dev & 0.47& 0.0& \cancel{0.0} & 0.06& \cancel{0.0} & 0.47& 0.0& \cancel{0.0} & 0.35  & 0.0 \\

\bottomrule
\end{tabular}}
\label{tab:ccms}
\end{table}



We also evaluate 10 distinct issue-inducing attributes' mean and Std Dev (standard deviation) of corresponding Precision values of configuration compatibility bugs, which are shown in Table \ref{tab:ccms} (- represents failing to detect issue-inducing XML elements). We remove attributes No. 3, No. 5, and No. 8 because they correspond to configuration compatibility bugs with only one occurrence. The non-canceled attributes are associated with corresponding configuration compatibility bugs, each of which has a count of at least 3 occurrences. From the table, we can observe that the mean values for the majority of \textit{<LLM, Attribute>} pairs are quite low, with only 3 pairs having a mean of 1.0, while the rest are all below 0.7. Furthermore, even for \textit{<LLM, Attribute>} pairs with mean values in the range of 0.5 to 0.7, their Std Dev values are relatively high (e.g., \textit{<Bard, No. 1>}'s 0.63 mean value and 0.47 Std Dev value), indicating the instability performance of LLMs' detection for these attributes. In general, LLMs perform poorly in detecting the seven types of issue-inducing attributes, and they almost do not exhibit strong detection capabilities for any of these attributes.

By manually analyzing these responses from LLMs, we find that the main reason for the low precision is two-fold: (1) the detection of LLMs is very sensitive. They always use 'may' to indicate many uncertain configuration compatibility bugs (but they may still miss issue-inducing attributes); and (2) LLMs may detect bugs under Android API level 21. Thus, we append the minimum constraint to the prompt and re-evaluate this RQ. After updating the prompt, we find that the performance of LLMs has no significant improvement. The Recall value in XEC of Bard drops to 76.6\% and even the one of GPT-4 drops to 20.8\% with 16.9\% Recall value in AC. Only GPT-3.5 improves slightly, reaching 80.5\%, 72.7\%, 24.2\%, and 41.3\% in the four metrics, respectively. However, the performance is still ineffective. Thus, based on all the results presented above, we can conclude that the configuration compatibility bug detection performance of LLMs is poor. They not only fail to effectively identify issue-inducing attributes but also can introduce a significant number of false positives.

\begin{tcolorbox}[boxrule=1pt,boxsep=1pt,left=2pt,right=2pt,top=2pt,bottom=2pt]
\textbf{Answer to RQ1:}
LLMs exhibit diverse configuration compatibility bug detection performance in XEC. However, their issue-inducing attribute detection capability is poor. They not only fail to effectively identify issue-inducing attributes but also can introduce a significant number. 
\end{tcolorbox} 


\subsection{RQ 2: Can LLMs identify conflicting Android API levels for compatibility bugs?}\label{subsec:rq2}

\noindent\textbf{Motivation.} Configuration compatibility bugs may occur across certain Android API levels. Thus, we intend to evaluate whether LLMs can identify the potential conflicting Android API levels.

\noindent\textbf{Methodology.} We assume that we already know there are configuration compatibility bugs in XML elements with corresponding attributes. This assumption is reasonable for only evaluating the capacity of LLMs identifying conflicting Android API levels. We provide bugs and neighboring API levels to LLMs to identify conflicting Android API levels. Specifically, we list out each set of potential conflicting API levels. With the minimum and maximum Android API levels set to 21 and 31, respectively, there are a total of 10 possible cases in $\{ (l_1, l_1 + 1) \mid 21 \leq l_1 \leq 30 \}$. For each case $(l_1, l_1 + 1)$, we construct prompts by combining them with bugs and input them to LLMs. We set the LLM task to whether issue-inducing XML elements and their corresponding issue-inducing attributes can cause configuration compatibility bugs between the two Android API levels $l_1$ and $l_1 + 1$. We also require LLMs only to consider these two API levels, preventing LLMs from considering other Android API levels. Compared to asking LLMs which two conflicting Android API levels exist, this task setting advantageously specifies and fixes the search space for conflicting Android API levels. We specify that LLMs need to answer [Yes] if the provided potential conflicting Android API level $(l_1, l_1 + 1)$ is the true one for the bug; otherwise, LLMs need to answer [No]. 


In this RQ, we utilize two metrics for evaluation: Recall and Precision. Recall represents the rate of truly identified conflicting Android API levels to the ground truth, and Precision represents the rate of truly identified conflicting Android API levels to all identified levels. Like Sec. \ref{subsec:rq1} (RQ 1), we set both GPT-3.5 and GPT-4's temperatures to 0 and query LLMs with 1 run for each prompt.




\begin{table}[t]
\centering
\caption{Identifying Conflicting Android API Levels Results by GPT-3.5, GPT-4, and Bard}
\scalebox{0.8}{\begin{tabular}{c|c|c|c|c|c|c} 
\toprule
\textbf{LLM} & \multicolumn{2}{c|}{GPT-3.5} & \multicolumn{2}{c|}{GPT-4} & \multicolumn{2}{c}{Bard}   \\ 

\hline
\textbf{Metric} & Recall & Precision  & Recall & Precision & Recall & Precision \\

\hline

\textbf{Value} & 42.9\% & 10.8\%  & 71.4\% & 16.0\% & 48.1\% & 11.5\% \\

\bottomrule
\end{tabular}}
\label{tab:ricaal}
\end{table}

\begin{figure}[t]
    \centering
    \subfigure{
    \includegraphics[width=0.32\textwidth]{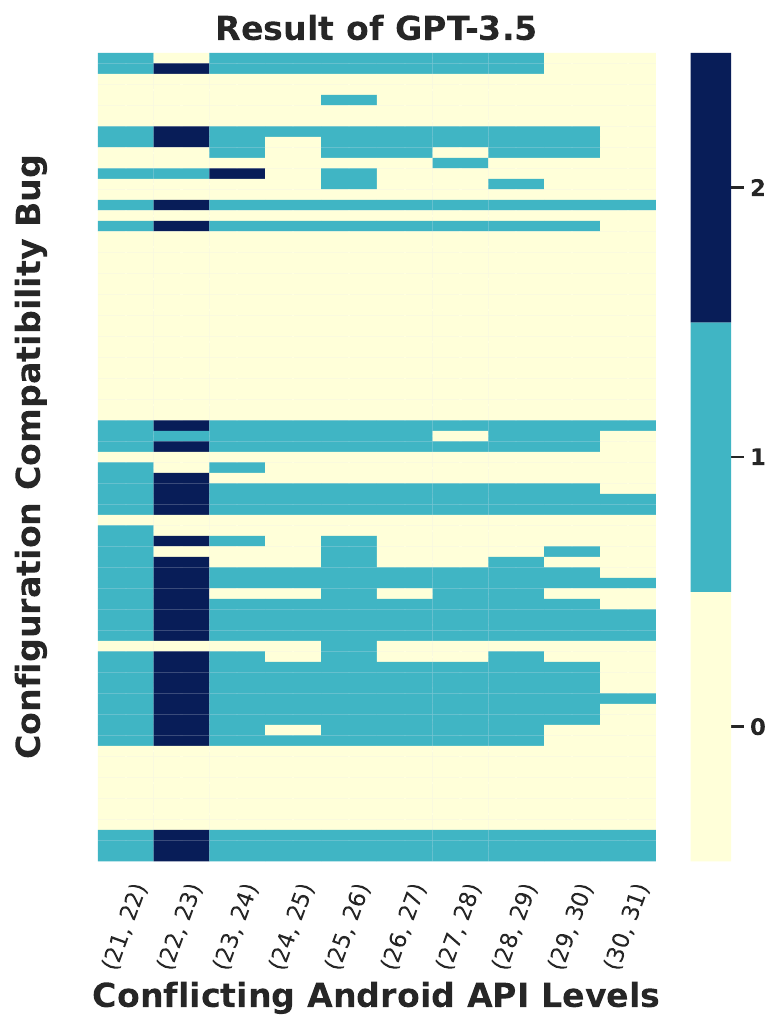}}
    \subfigure{
    \includegraphics[width=0.32\textwidth]{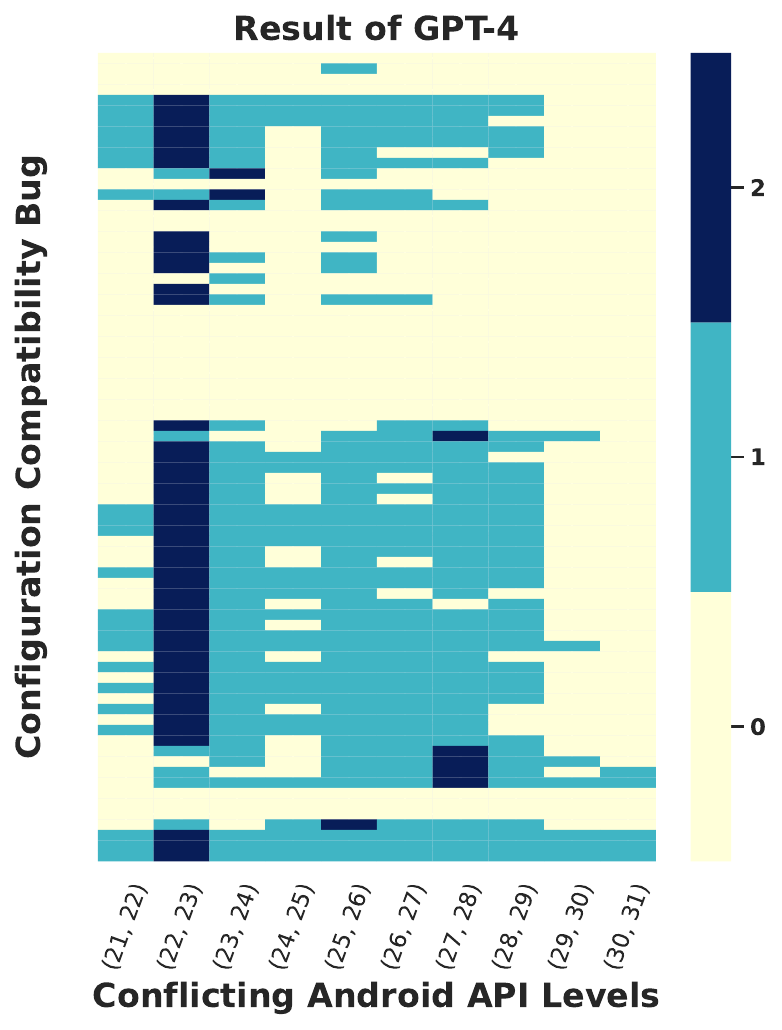}}
    \subfigure{
    \includegraphics[width=0.32\textwidth]{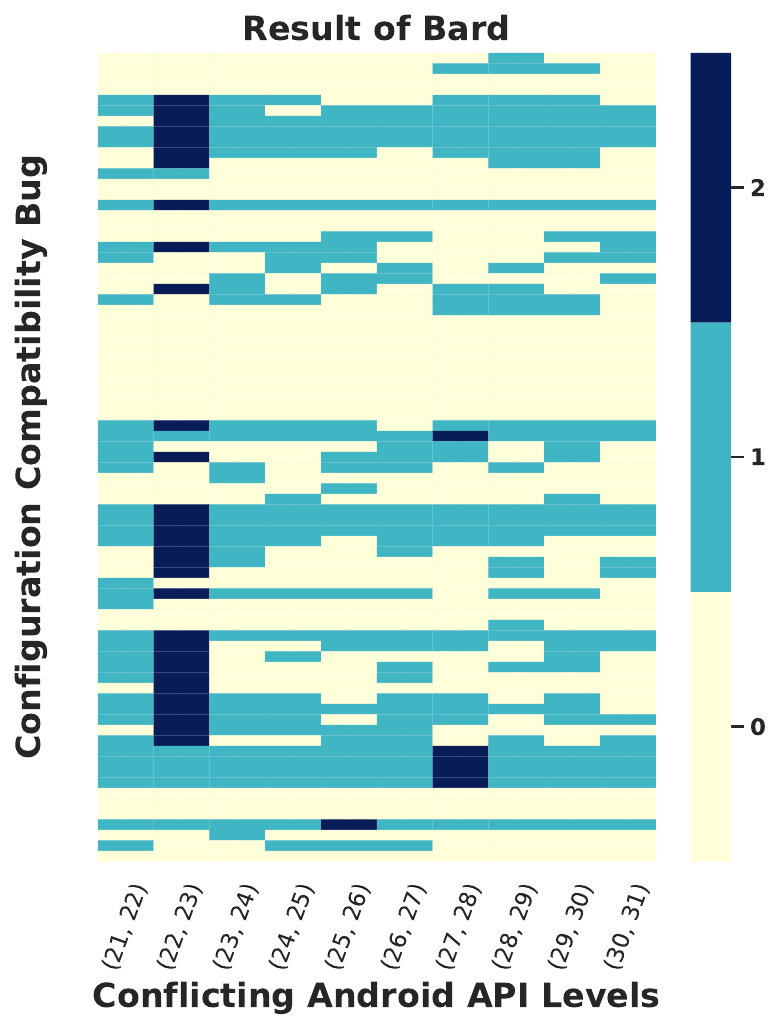}}
    \caption{Results in heatmaps of identifying conflicting Android API levels by GPT-3.5, GPT-4, and Bard.}
    \label{fig:heatmap-conflict}
\end{figure}

\noindent\textbf{Results.} The experimental results are shown in Table \ref{tab:ricaal}. Recall values of three LLMs are all below 72\%, with 42.9\%, 71.4\%, and 48.1\% to GPT-3.5, GPT-4, and Bard, respectively. The results indicate that LLMs have poor performance in identifying conflicting Android API levels. Moreover, the Precision values for all three LLMs are very low, only with 10.8\%, 16.0\%, and 11.5\% to GPT-3.5, GPT-4, and Bard, respectively, indicating that LLMs can introduce a significant number of false positives of the identified conflicting Android API levels. To get a better intuition for the results, we show the identified results in three heatmaps (see Fig. \ref{fig:heatmap-conflict}) for GPT-3.5, GPT-4, and Bard, where the y-label represents each configuration compatibility bug and the discrete numbers 0, 1, and 2 floating to the right of color bars represent unidentified, truly identified, and falsely identified conflicting Android API levels, respectively, for each bug. From the figure, we can observe that LLMs have poor identifying performance. Regardless of GPT-3.5 and Bard, they all have many rows in the figure that are shaded with light yellow (i.e., no true and false conflicting Android API level is found). For GPT-4, it still has many same condition rows but its Recall values are higher (Table. \ref{tab:ricaal}). Furthermore, for the rows where conflicting Android API levels are identified, we can also observe a significant number of false positives, and in all cases, the deep blue cells are surrounded by light blue cells on one or both sides within the same row, for all three LLMs. By manually analyzing the responses from LLMs, we find that they are unaware of some issue-inducing attributes and consider Android API levels out of the bounds of conflicting Android API levels though we add a constraint in prompt, causing mediocre Recall and low Precision, respectively. Thus, in general, based on these results, we can conclude that LLMs cannot effectively identify conflicting Android API levels for configuration compatibility bugs.


\begin{tcolorbox}[boxrule=1pt,boxsep=1pt,left=2pt,right=2pt,top=2pt,bottom=2pt]
\textbf{Answer to RQ2:}
LLMs are ineffective in identifying conflicting Android API levels for configuration compatibility bugs. Specifically, they are unable to identify conflicting Android API levels accurately and also introduce a significant number of false positives.
\end{tcolorbox}

\subsection{RQ 3: How well do LLMs perform at repairing configuration compatibility bugs?}\label{subsec:rq3}

\noindent\textbf{Motivation.} Furthermore, we are interested in whether LLMs can be used to repair these bugs. Therefore, we evaluate the performance of LLMs in repairing configuration compatibility bugs. 

\noindent\textbf{Methodology.} To answer this research question, we assume that we already know the complete configuration compatibility bug information. We provide the complete bug information to LLMs to repair bugs. The information given includes issue-inducing XML elements, issue-inducing attributes, and conflicting Android API levels. Moreover, to standardize the repairing process of LLMs, we follow the guideline of code maintenance~\cite{ouni:2016-multi}: \textit{provide as few modifications as possible}. Specifically, we set 6 repairing rules combined together in one prompt template for LLMs based on the guideline, \cite{Huang:2021} and \cite{Huang:2023} which are listed as follows:

    \noindent $\bullet$ \textbf{Rule 1:} Attempt to repair bugs by modifying the issue-inducing attributes or their values as little as possible. The modifying approaches include (1) specifying issue-fixing attributes in issue-inducing XML elements to eliminate inconsistent runtime behavior, (2) changing the values of issue-inducing attributes, and (3) removing issue-inducing attributes. This rule follows \cite{Huang:2023}.

    \noindent $\bullet$ \textbf{Rule 2:} If \textbf{rule 1} is inapplicable, then specify other different types of XML elements to replace the issue-inducing XML elements.

    \noindent $\bullet$ \textbf{Rule 3:} Ensure the repairing changes to XML elements are minimized as much as possible.

    \noindent $\bullet$ \textbf{Rule 4:} LLMs cannot introduce Android API calls or create new resource files or folders in the repaired results. This rule also follows \cite{Huang:2023}.

    \noindent $\bullet$ \textbf{Rule 5:} The visual effects of the repaired apps are consistent between the two conflicting Android API levels. 

    \noindent $\bullet$ \textbf{Rule 6:} The visual effects of the repaired apps should be as similar as possible to the original apps before repairing on Android API level $l_1 + 1$. The last two rules tell LLMs the repairing goal.

We set the LLM task to repair bugs by using complete bug information and the above 6 rules. We also specify that LLMs need to respond with the complete repaired XML elements.

We utilize three metrics by following \cite{Huang:2023} to evaluate the repairing performance of LLMs: \textit{Correct}, \textit{Overfitting}, and \textit{Correct@k}. We define the original apps with configuration compatibility bugs as \textsf{app} and the repaired apps as \textsf{app$_{rc}$}. Then, \textit{Correct} represents \textsf{app$_{rc}$} has the same runtime behavior (i.e., visual effects) at Android API level $l_t = 31$ as \textsf{app} with eliminating inconsistencies at API level $l_1$, and introduces no other unintended runtime behavior (e.g., crashes) due to repairing. \textit{Overfitting} means that the repairing process introduces issue-fixing attributes but also causes unintended runtime behavior (e.g., introduce other unnecessary XML elements causing crashes). \textit{Correct@k} evaluates the success rate of LLMs in configuration compatibility bug repairing. A bug is considered repaired successfully if at least one of \textit{k} patches (repairs) generated by LLMs can repair the bug. \textit{k} is set to 5, consistent with the setting used in \cite{Huang:2023}. It is important to note that the same visual effects need to be assessed manually, which introduces subjectivity. Therefore, we have two graduate students studying computer science perform the assessment together to minimize the impact of subjectivity. We also set both GPT-3.5 and GPT-4's temperatures to 0.7 to strengthen their randomness and creativity. This setting is the default option in OpenAI API document~\cite{OpenAI-API}.

\begin{table}[t]
\centering
\caption{Repairing Results for Configuration Compatibility Bugs by GPT-3.5, GPT-4, and Bard}
\scalebox{0.8}{\begin{tabular}{c|c|c|c|c|c|c|c|c|c} 
\toprule
\textbf{LLM} & \multicolumn{3}{c|}{GPT-3.5} & \multicolumn{3}{c|}{GPT-4} & \multicolumn{3}{c}{Bard}   \\ 

\hline
\textbf{Metric} & \textit{Correct} & \textit{Overfitting} & \textit{Correct@k}  & \textit{Correct} & \textit{Overfitting} & \textit{Correct@k} & \textit{Correct} & \textit{Overfitting} & \textit{Correct@k} \\

\hline

\textbf{Value} & 15.3\% & 0.3\%  & 31.2\% & 53.8\% & 1.6\% & 74.0\% & 2.5\% & 0.0\% & 5.2\% \\

\bottomrule
\end{tabular}}
\label{tab:repairing-result-rq3}
\end{table}

\noindent\textbf{Results.} The repairing results of three LLMs are shown in Table \ref{tab:repairing-result-rq3}. The performance for repairing bugs by LLMs is not good. Bard performs the worst, only achieving 2.5\% \textit{Correct} value and 5.2\% \textit{Correct@k} value. GPT-3.5 can also only achieve 15.3\% \textit{Correct} value and 31.2\% \textit{Correct@k} value. GPT-4 performs the best but still with slightly low results of 53.8\% \textit{Correct} value and 74.0\% \textit{Correct@k} value. The gap between \textit{Correct} and \textit{Correct@k} also indicates that the randomness of LLMs makes the stability of repairing weak. The \textit{Overfitting} values of them are all zero or close to zero, indicating that the reason for the failures of repairing is that LLMs have difficulty finding suitable patches.

\begin{table}[t]
\centering
\caption{Failure Categories in All Failed Repairs for GPT-3.5, GPT-4, and Bard}
\scalebox{0.7}{\begin{tabular}{clccc}
\hline
\multicolumn{1}{c}{\multirow{2}{*}{\textbf{Category}}} & \multicolumn{1}{c}{\multirow{2}{*}{\textbf{Explaination}}} & \multicolumn{3}{c}{\textbf{Frequency}} \\ \cline{3-5}
                       &       &  \textbf{GPT-3.5}  & \textbf{GPT-4} & \textbf{Bard}  \\ \hline
\textsf{C-R} & Fail to link file resources when compiling. & 27  &  14 & \textcolor{red}{98} \\
\textsf{C-A} & Fail to find attributes when compiling. & 7   &  0 & \textcolor{red}{125} \\
\textsf{U-I} & \multicolumn{1}{m{13cm}}{Update values of issue-inducing attributes (and introduce issue-unrelated attributes or elements).} &  17   &  10  & 59 \\
\textsf{R-I} & \multicolumn{1}{m{13cm}}{Remove issue-inducing attributes (and introduce issue-unrelated attributes or elements).} & \textcolor{red}{246}   &  \textcolor{red}{113}  &  72  \\
\textsf{I-A} & \multicolumn{1}{m{13cm}}{Introduce issue-fixing attributes but append other issue-unrelated attributes or elements, causing failures.} & 0  &  6  &  0 \\
\textsf{N-I} & \multicolumn{1}{m{13cm}}{No repairs for issue-inducing attributes (and introduce issue-unrelated attributes or elements).} & 25  & 26 & 29 \\ \hline
\end{tabular}}
\label{tab:categories-of-failures}
\end{table}

Furthermore, we manually analyze these three LLMs' results and find that all failed repairs can be classified into 6 categories as shown in Table \ref{tab:categories-of-failures}. \textsf{C-R} and \textsf{C-A} are related to compile errors, caused by non-existent resources and attributes (e.g., \code{android:drawableTintCompat}). GPT-3.5 and GPT-4 have relatively few errors in them, with 34 and 14, respectively. They have a low probability of introducing non-existent resources and attributes. However, for Bard, it has a lot with 223 in total, and \textsf{C-R} and \textsf{C-A} account for 58.5\% among all failures. Through a deep analysis for attributes and values introduced by Bard, we find that almost all errors originate from typos (e.g., change \code{layout\_constraintStart\_toEndOf} to \code{layout\_constraintStart\_toEnd}) and inappropriate new resources (e.g., replace \code{@drawable/ic\_logo\_mail} with \code{@drawable/ic\_logo\_mail\_24}). This result indicates that Bard is ineffective in guaranteeing consistency between input XML elements and output ones. \textsf{U-I} is related to updating values of issue-inducing attributes. GPT-3.5 and GPT-4 also have few errors in this category. Bard still has many failures in this category with 59 of them. LLMs update the values but compatibility bugs are caused by issue-inducing attributes themselves. For example, \code{"center"} for \code{android:gravity} is changed to \code{"center\_vertical|center\_horizontal"} but the bug is not eliminated between conflicting Android API levels 22 and 23 since the attribute is introduced in Android API level 23~\cite{Android-Developers}. \textsf{R-I} is related to removing issue-inducing attributes. GPT-3.5, GPT-4, and Bard all have many failures in this category, with 246, 113, and 72, respectively. These three LLMs perform this operation so that compatibility bugs are eliminated, but the repaired apps have different runtime behavior (e.g., visual effects) compared with corresponding unrepaired ones. \textsf{I-A} is related to \textit{Overfitting} that issue-fixing attributes are introduced but some additional issue-unrelated attributes are also appended, causing failures such as crashes. GPT-4 has 6 failures in this category, and GPT-3.5 and Bard have no failures. One example is that GPT-4 changes <\code{TextView}> to <\code{android.support.v7.widget.AppCompatTextView}> and causes a crash due to a library dependency issue (i.e., \code{ClassNotFoundException}). \textsf{N-I} is related to deceptive repair, which means no repairing is made. GPT-3.5, GPT-4, and Bard have a few failures in this category. In the repairing process, they respond issue-inducing attributes and values cause no compatibility bug. 

We also observe the cases of \textit{Correct} and failures by GPT-3.5 and GPT-4 (we ignore Bard due to its quite low metrics) and find that these two LLMs can repair (hard-to-repair) bugs that are hard or impossible to repair by traditional tools (e.g., ConfFix~\cite{Huang:2023}). The definition of a hard-to-repair bug is it cannot be repaired in 100\% probability by a pre-defined repairing rule through ConfFix~\cite{Huang:2023}. The rules for repairing it need to be adjusted dynamically (e.g., positions) or go beyond known rules. We choose ConfFix as the reference because it is the most comprehensive known tool for repairing rules. One example is repairing for \code{android:layout\_height="match\_parent"} that the value is changed to \code{"wrap\_content"}, which is impossible for traditional tools due to their limitation on searching for nonnumerical values~\cite{Huang:2023}. Additionally, another example already is shown in Fig. \ref{fig:example-compatibility-bug-intro} (see Sec. \ref{sec:motivationexample}) that traditional tools have limited repairing rules to the bug related to <\code{ImageView}> with \code{android:foreground}. Both GPT-3.5 and GPT-4 can successfully repair it by introducing additional XML element <\code{FrameLayout}> or issue-fixing attribute \code{android:background}, which are inapplicable for traditional tools. However, these two LLMs perform poorly in repairing relatively easy-to-repair bugs (e.g., directly replacing issue-inducing attributes and maintaining original values), the opposite of hard-to-repair bugs. GPT-3.5 and GPT-4 fail to produce 43 and 14 correct repairs for such bugs, respectively.

Therefore, we can conclude that GPT-3.5 and GPT-4 have a potential capability for repairing compatibility bugs, especially for the bugs that are impossible to be repaired by traditional tools.

\begin{tcolorbox}[boxrule=1pt,boxsep=1pt,left=2pt,right=2pt,top=2pt,bottom=2pt]
\textbf{Answer to RQ3:}
In general, the repairing capabilities of GPT-3.5 and GPT-4 are weak and unstable. GPT-3.5 achieves 15.3\% \textit{Correct} value and 31.2\% \textit{Correct@k} value and GPT-4 achieves 53.8\% and 74.0\%. The failures of repaired results mainly drop in \textsf{R-I} category, and the repairing of easy-to-repair bugs is poor. However, these two LLMs can repair bugs that are hard to repair by traditional tools. Bard is ineffective in repairing configuration compatibility bugs.
\end{tcolorbox}

\subsection{Insights Found}\label{sec:d-ll}

\noindent \textbf{Insight 1.} Through the results from Sec. \ref{subsec:rq1} and Sec. \ref{subsec:rq2}, we conclude that LLMs are not effective in performing configuration compatibility bug detection. The possible reason is that there is less bug-relevant data in the training dataset. For example, there are only two search results for the \textit{seekbar} bug~\cite{Neko:885c7bb} (i.e., use \code{"wrap\_content"} for \code{android:layout\_height} instead of \code{"match\_parent"} in seek bar-related XML element) between conflicting Android API levels 22 and 23 on top of Stack Overflow~\cite{stackoverflow}, and only one of them has an answer (both of them are before 2021). These results indicate LLMs not being able to learn to recognize why configuration compatibility bugs arise~\cite{carlini2021extracting}. 

\noindent \textbf{Insight 2.} However, LLMs perform better in configuration compatibility bug repairing, though in general, their capabilities are weak and unstable. In particular, LLMs show great repairing flexibility for hard-to-repair bugs. We believe that LLMs are able to show performance in the repairing task because repairing is considered to be actually a prediction task for LLMs. The input prompt tells LLMs the location and cause of the bug, so all they need to do is to make predictions based on that information to repair the bug. LLMs learn a lot of Android configuration XML code on the internet (e.g., GitHub), thus they are also able to predict corresponding repaired XML code~\cite{carlini2021extracting}. Hence, we believe that a prompt with more explicit repair suggestions may potentially enhance the performance of the LLMs in repairing configuration compatibility bugs. This is premised on the idea that a more detailed prompt may provide LLMs with a clearer understanding of bugs, enabling them to make more accurate predictions for the repairs~\cite{liu:2023-chatting, pearce:2023, liu2023refining}.

\section{LLM-CompDroid}\label{sec:Approach}

\begin{figure}[t]
    \centering
    \includegraphics[width=1.0\textwidth]{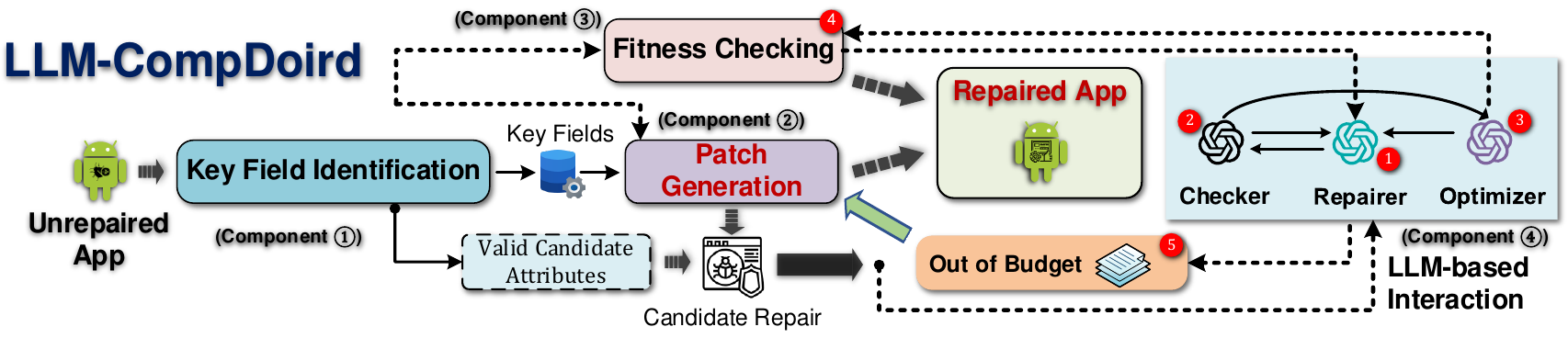}
    \caption{Overview of LLM-CompDroid framework.}
    \label{fig:workflow}
\end{figure}

\subsection{Overview of LLM-CompDroid}
Though LLMs show performance in repairing hard-to-repair bugs, in general, their repairing capabilities are weak and unstable. In this section, we introduce the overall design of our LLM-CompDroid framework for repairing configuration compatibility bugs by leveraging the advantages from both LLMs and traditional tools. Moreover, in this study, we leverage ConfFix~\cite{Huang:2023}, a dynamic analysis repairing tool, as the representative of traditional tools due to that it is the most comprehensive known tool for repairing rules, where its repair rules are more general compared to other tools~\cite{Lint, XFix}. As illustrated in Fig. \ref{fig:workflow}, the LLM-CompDroid contains the following components: \whiteding{1} Key Field Identification, \whiteding{2} Patch Generation, \whiteding{3} Fitness Checking, and \whiteding{4} LLM-based Interaction. Given a configuration compatibility bug, LLM-CompDroid first leverages component \whiteding{1} - \whiteding{3} from ConfFix~\cite{Huang:2023} to identify key fields $F$ affected by issue-inducing attributes $A_b$ and generate a single one-line patch (i.e., replace $A_b$ with one issue-fixing attribute $A_f$ or remove $A_b$) to repair the bug. If the (easy-to-repair) bug is repaired directly, then LLM-CompDroid stops; otherwise, LLM-CompDroid startups component \whiteding{4} to perform further LLM-based repairing by leveraging LLMs' advantages and integrate component \whiteding{3} as a fitness function to guide the subsequent process for repairing the (hard-to-repair) bug. If component \whiteding{4} still cannot find a possibly correct patch, then LLM-CompDroid uses component \whiteding{2} to generate a combination patch as the final repaired result. 

\subsection{Repairing Configuration Compatibility Bugs}

Here, we first introduce ConfFix~\cite{Huang:2023} APIs (i.e., component \whiteding{1}, \whiteding{2}, and \whiteding{3}) used in LLM-CompDroid. Then we demonstrate the component \whiteding{4} LLM-based Interaction including three agents. Finally, we present the interactions among these agents and ConfFix APIs.

\subsubsection{ConfFix APIs}

\noindent 

ConfFix APIs serve as the base of LLM-CompDroid, repairing easy-to-repair bugs and providing additional repairing information (candidate repair) to LLM-based Interaction for hard-to-repair bugs to possibly shrink down the search space of patches. 

\noindent \textbf{Key Field Identification.} Key fields $F$ are the ones holding inconsistent values across conflicting Android API levels due to configuration compatibility bugs. Thus, with the help of $F$, the problem of repairing bugs can be transformed into monitoring $F$'s runtime status. Therefore, component \whiteding{1} is the base of subsequent components \whiteding{2} - \whiteding{4}. Specifically, ConfFix identifies key objects $O$ related to the issue-inducing XML element by leveraging instrumentation based on corresponding resource \code{id}. After identifying $O$, ConfFix can monitor the fields in $O$. Thus, ConfFix removes the issue-inducing attributes $A_b$ from the unrepaired \textsf{app} to generate an \textsf{app$'$} and runs both of them at a high Android API level (e.g., 31) to identify fields with different values as $F$ by using access paths~\cite{tripp2013andromeda}. We also output valid candidate attributes (e.g., \code{android:top} to \code{android:gravity}) here for replacing $A_b$ according to the empirical rules in ConfFix in this component since ConfFix may fail to identify $O$ in some cases~\cite{Huang:2023}. Without $F$, component \whiteding{2} and \whiteding{3} are out of work and thus, these candidate attributes can still be provided to LLMs to possibly shrink down the search space of patches. 

\noindent \textbf{Patch Generation.} After identifying $F$, ConfFix replaces $A_b$ 
 with each candidate attribute to generate multiple \textsf{app$_r$} with single one-line patches. Then, ConfFix tests each \textsf{app$_r$}, records $F$'s values, and compares these $F$'s values to the values in \textsf{app} and \textsf{app$'$} by leveraging fitness function (i.e., component \whiteding{3}). This process continuously adjusts the value of each candidate attribute within a budget to improve the score from the fitness function, in order to find a value for each candidate attribute maximizing the score. The fitness function evaluates a score ranging from $-\infty$ to 1. The score represents the normalized difference between \textsf{app$_r$} and \textsf{app}'s $F$, at low and high Android API levels (across conflicting Android API levels), respectively. A higher value indicates a smaller difference, and 1 represents no difference. If there exists an \textsf{app$_r$} with a score of 1, then \textsf{app$_r$} is considered a possibly correct patch (patching strategy 1); otherwise, ConfFix considers the score of each \textsf{app$_r$} as a probability and samples the corresponding single one-line patches to generate multiple combined patches (\textsf{app$_c$}) from sampled single one-line patches (i.e., one combined patch is generated from combining the corresponding sampled single one-line patches). It also uses the fitness function to find the best patch among these combined patches as the final repaired result (patching strategy 2).

\noindent \textbf{Fitness Checking.} The fitness function evaluates the difference between repaired \textsf{app$_{rc}$} (i.e., \textsf{app$_r$} and \textsf{app$_c$}) and \textsf{app}. It is shown in the following equation \ref{equ:fitness}:

\begin{equation}
   Score(\text{\textsf{app}}_{rc}) = 1 - \frac{FDiff(\text{\textsf{app}}_{rc}, \text{\textsf{app}})}{FDiff(\text{\textsf{app$'$}}, \text{\textsf{app}})} \label{equ:fitness}
\end{equation}

Where function $FDiff$ measures the difference of $F$'s values between the input two apps across conflicting Android API levels or at a high Android API level\footnote{ConfFix evaluates $FDiff$ only at a high Android API level. However, LLMs may repair nothing for issue-inducing XML elements, thus we also introduce evaluating at conflicting Android API levels to avoid LLMs' deceptive repairs.}, ranging from 0 to $+\infty$. $FDiff(\text{\textsf{app$'$}}, \text{\textsf{app}})$ is used for normalization, at a high Android API level. If the evaluated score is not equal to 1, then Fitness Checking outputs fitness-based feedback information (e.g., see Sec. \ref{subsec:req4}) for subsequent repairing. If the fitness function is inapplicable, including failing to identify $F$ or introducing a different XML element to replace or wrap the issue-inducing XML element, then Fitness Checking considers the repaired result as a possibly correct patch directly.

The details of these three components can be found here~\cite{Huang:2023}.

\subsubsection{LLM-based Interaction}

\noindent

LLM-based interaction leverages the collaboration of three LLM-powered agents: Repairer, Checker, and Optimizer. Each agent maintains its own session with corresponding conversations and context. Additionally, these three agents focus on repairing-related tasks. Thus, through the interaction between the three agents, LLM-CompDroid converges to a final decision of patch for repairing configuration compatibility bugs.

\noindent \textbf{Repairer.} The goal of Repairer is to generate a patch for an input prompt $x$ with a configuration compatibility bug. Given an $x$, Repairer generates the corresponding repaired XML element with explanation. There are two types of $x$, $x_i$ and $x_c$. $x_i$ is the initial prompt sent to Repairer, containing the background information on configuration compatibility bugs with repairing rules (see Sec. \ref{subsec:rq3}), the specific configuration compatibility bug, the task of repairing, and the potential candidate repair information (PCRI) from component \whiteding{1} or \whiteding{2}. $x_c$ is the continuous prompt, which is feedback from Checker, Optimizer, or Fitness Checking. Thus, the whole conversation in Repairer contains three types of information, $x_i$, $x_c$, and Repairer's responses $r_r$ containing the repaired result and corresponding explanation, in order. Repairer is the only agent that generates repaired results. 

\noindent \textbf{Checker.} Repairer may generate repaired results with errors, so there is a need for additional mechanisms to detect these errors. As the checking agent, Checker is responsible for detecting whether errors (e.g., the introduced attribute is unable to repair bugs or has compile errors) exist in the patch generated by Repairer. The advantages of Checker in this task is that an LLM-powered agent is trained on an extensive dataset and thus, can supply more flexible and informative feedback to Repairer than traditional rule-based tools, especially for hard-to-repair bugs where ConfFix can only provide candidate attribute information. Specifically, Checker receives two types of prompts, $y_i$ and $y_c$. $y_i$ only provides the information of the background and Checker's task and $y_c$ includes the given configuration compatibility bug and the last response $r_r$ from Repairer. Moreover, Checker responds the checking result $r_c$, one type of $x_c$, with either an error-free or error-related explanation. 

\noindent \textbf{Optimizer.} Though Repairer can generate correct patches for bugs, the repaired results may be more complex (e.g., introduce many issue-unrelated attributes or elements) than the original unrepaired XML elements, which can make it difficult for users to locate issue-fixing attributes or elements. Such examples can be found in Sec. \ref{subsec:rq3}. Optimizer determines whether the patch generated by Repairer can be further optimized (e.g., remove introduced issue-unrelated attributes). The advantages of Optimizer are the same as Checker and traditional tools cannot provide any valuable information in this aspect. Optimizer also receives two types of prompts, $z_i$ and $z_c$. Like Checker, $z_i$ provides the information of the background and Optimizer's task. $z_c$ includes the given configuration compatibility bug and the last response $r_r$, and requires Optimizer to check if the repaired results follow the repairing rules (see Sec. \ref{subsec:rq3}). Moreover, Optimizer responds the checking result $r_o$, one type of $x_c$, with either an optimization-free or optimization-related explanation.

The prompts used in these three agents follow the principle of prompt design.

\subsubsection{Workflow}

\noindent

The specific workflow of LLM-CompDroid (see Fig. \ref{fig:workflow}) is shown as follows:

\noindent \textbf{Step \blackding{1}.} Given a configuration compatibility bug $b$ to the unrepaired \textsf{app}, LLM-CompDroid first utilizes component \whiteding{1} to identify key fields $F$ in \textsf{app}. If $F$ are identified, then component \whiteding{1} outputs $F$ and valid candidate attributes to component \whiteding{2} (next step: Step \blackding{2}); otherwise, it sends the candidate attributes as PCRI to component \whiteding{4} (next step: Step \blackding{3}) since the fitness function is inapplicable. 

\noindent \textbf{Step \blackding{2}.} In this step, component \whiteding{2} receives the outputs in Step \blackding{1} and uses them to resolve the single one-line patch of each candidate attribute based on component \whiteding{3}. If there exists a patch satisfying patching strategy 1, then LLM-CompDroid outputs the repaired \textsf{app$_r$} and stops; otherwise, component \whiteding{2} outputs all single one-line patches as PCRI to component \whiteding{4} (next step: Step \blackding{3}).

\noindent \textbf{Step \blackding{3}.} After receiving the PCRI from previous steps, LLM-CompDroid first constructs prompt $x_i$ to Repairer. Repairer receives $x_i$ and starts component \whiteding{4}, LLM-based Interaction, to converge a patch for repairing the bug $b$. Specifically, \textcolor{red}{\blackding{1}} Repairer receives a prompt $x_i$ (only for the first round) or $x_c$ to generate response $r_r$; \textcolor{red}{\blackding{2}} Checker ($y_i$ is set as system content~\cite{OpenAI-API}. $z_i$ is the same to Optimizer) receives $r_r$ and integrates it with the bug $b$ as the prompt $y_c$. If Checker determines there is no error, then it passes $r_r$ to Optimizer in \textcolor{red}{\blackding{3}}; otherwise, Checker sends its response $r_c$ to \textcolor{red}{\blackding{1}} as the prompt $x_c$; \textcolor{red}{\blackding{3}} Optimizer takes in $r_r$ and the bug $b$ as $z_c$ and generate a response $r_o$. If Optimizer determines there is no optimization space, then it extracts the repaired result and sends it to component \whiteding{3} in \textcolor{red}{\blackding{4}}; otherwise, Optimizer sends its response $r_0$ to \textcolor{red}{\blackding{1}} as the prompt $x_c$; \textcolor{red}{\blackding{4}} Fitness Checking evaluates the repaired result. If it is a possibly correct patch, LLM-CompDroid generates the repaired \textsf{app$_r$}; otherwise, Fitness Checking sends fitness-based feedback information back to \textcolor{red}{\blackding{1}} as the prompt $x_c$. The entire iterative interaction loop from \textcolor{red}{\blackding{1}} to \textcolor{red}{\blackding{2}}-\textcolor{red}{\blackding{4}} continues $n$ times (multi-round budget) and we count one loop as \textcolor{red}{\blackding{1}} being the starting point. If Repairer generates the same repaired result \textsf{app$_r$} twice consecutively, with the loop going through \textcolor{red}{\blackding{4}}, LLM-CompDroid considers it a possibly correct patch. Moreover, if component \whiteding{4} cannot generate an \textsf{app$_r$} (e.g., continuous compile errors), then LLM-CompDroid performs the step \blackding{4} to \textcolor{red}{\blackding{5}}.


\noindent \textbf{Step \blackding{4}.} \textcolor{red}{\blackding{5}} LLM-CompDroid goes back to component \whiteding{2} and leverages patching strategy 2 to generate an \textsf{app$_c$} as the final repaired result.

\section{Evaluation}\label{sec:Evaluation}

\subsection{Evaluation Setup}

This section introduces the setup for evaluating LLM-CompDroid.

\noindent \textbf{Dataset:} In this experiment, we use the same dataset introduced in Sec. \ref{sec:dataset-setup}.

\noindent \textbf{Experimental Setups:} The applicable maximum and minimum Android API level is set to 31 and 21, respectively. GPT-3.5 and GPT-4 are selected as the representative LLMs for evaluation. We do not choose Bard because it is ineffective in repairing configuration compatibility bugs and has consistency problems between input XML elements and output ones (Sec. \ref{subsec:rq3}). Moreover, we carefully perform manual analysis to check the repaired results by running them dynamically. 

\noindent \textbf{Principle of Prompt Design:} The prompt design for LLM-CompDroid follows the same principle introduced in Sec. \ref{sec:dataset-setup} and almost the same settings in Sec. \ref{subsec:rq3}. The specific prompt template can be found in our artifact~\cite{artifact}. 


\noindent \textbf{Baseline Method:} We use two state-of-the-art traditional tools as our baseline methods:  (1) Lint~\cite{Lint} is a rule-based static analyzer in Android Studio~\cite{Android-Developers}. It has a set of rules for repairing configuration compatibility bugs and (2) ConfFix~\cite{Huang:2023} is a state-of-the-art dynamic analysis tool for repairing configuration compatibility bugs. It generates patches by leveraging pre-defined rules and monitoring the runtime status of key fields in the Android Framework across conflicting Android API levels. We also leverage pure GPT-3.5 and GPT-4 as baseline methods (see Sec. \ref{subsec:rq3}). In addition, we introduce LLM-after-ConFix which consists of trivially querying LLMs when ConFix fails to repair a bug. This comparison is important to assess the effectiveness of the combination of ConFix and LLMs (i.e., LLM-CompDroid), beyond merely using LLMs when ConFix is not enough.

\noindent \textbf{Environment:} We use the same environment introduced in Sec. \ref{sec:dataset-setup}.

\subsection{Research Questions}
The evaluation of LLM-CompDroid is led by answering the following research questions:

\noindent$\bullet$ \textbf{RQ 4\footnote{Here, we use bugs to refer to configuration compatibility bugs.}:} \textbf{How well do LLM-CompDroid perform at repairing bugs?}

\noindent$\bullet$ \textbf{RQ 5: How does LLM-CompDroid compare to traditional tools in repairing bugs?}

\subsection{RQ4: How well do LLM-CompDroid perform at repairing bugs?}\label{subsec:req4}

\noindent\textbf{Motivation.} In this RQ, we evaluate the performance of LLM-CompDroid in repairing configuration compatibility bugs. 

\noindent\textbf{Methodology.} The temperatures of GPT-3.5 and GPT-4 are set to 0.7 and the multi-round budget and time budget are set to 10 and 120 minutes. The time budget is consistent with existing practice~\cite{Huang:2023}. In addition, \textit{k} is set to 5 (due to LLMs' randomness) to be consistent with the setting used in~\cite{Huang:2023}, which means that for each bug, we run LLM-CompDroid to repair it 5 times. In total, there are 77 configuration compatibility bugs with 385 repairing times. The fitness-based feedback information is set with two options: \blackding{1} coarse-granularity general information (i.e., tell LLMs is there a better repair) and \blackding{2} fine-granularity difference information of key fields between the repaired app and the unrepaired one across conflicting Android API levels. If the repaired results cannot be compiled, then the fitness-based feedback information only includes the compile error information. We utilize three metrics: \textit{Correct}, \textit{Overfitting}, and \textit{Correct@k} for evaluation. 

\begin{table}[t]

\centering
\caption{Repairing Results for Configuration Compatibility Bugs by LLM-CompDroid and Baseline Methods}
\scalebox{0.9}{\begin{tabular}{c|c|c|c} 

\toprule

\textbf{Method} & \textbf{\textit{Correct}}  & \textbf{\textit{Overfitting}}   &  \textbf{\textit{Correct@k}} \\ 

\hline

LLM-CompDroid-GPT-3.5 \blackding{1} &  91.9\% & 1.6\%   & 93.5\%  \\ 
\hline
LLM-CompDroid-GPT-4 \blackding{1} &  \textbf{92.5\%} &  2.6\%  & \textbf{96.1\%}  \\ 
\hline
LLM-CompDroid-GPT-3.5 \blackding{2} &  92.2\% & 1.0\%   & 93.5\%  \\ 
\hline
LLM-CompDroid-GPT-4 \blackding{2} &  90.9\% &  1.0\%  & 93.5\%  \\ 
\hline
LLM-after-ConfFix-GPT-3.5  & 87.5\%  &  0.3\%  & 94.8\%  \\ 
\hline
LLM-after-ConfFix-GPT-4  & 89.4\%  &  0\%  & 94.8\% \\ 
\hline
GPT-3.5  & 15.3\%  & 0.3\%   & 31.2\% \\ 
\hline
GPT-4  &  53.8\% & 1.6\%   & 74.0\% \\ 
\hline
Lint &  41.6\% &  \textbf{0.0\%}  & - \\ 
\hline
ConfFix & 82.1\%  &  1.0\%  & 83.1\% \\ 
\bottomrule
\end{tabular}}
\label{tab:repairing-result-rq4}
\end{table}


\begin{table}[t]
\centering
\caption{Failure Categories in All Failed Repairs for LLM-CompDroid-GPT-3.5 and LLM-CompDroid-GPT-4}
\scalebox{0.9}{\begin{tabular}{ccc}
\hline
\multicolumn{1}{c}{\multirow{2}{*}{\textbf{Category}}}  & \multicolumn{2}{c}{\textbf{Frequency}} \\ \cline{2-3}
                     &    \textbf{LLM-CompDroid-GPT-3.5}  & \textbf{LLM-CompDroid-GPT-4}  \\ \hline
                     \textsf{C-R}  &  0   &  0   \\
                     \textsf{C-A}  &  0   &  0   \\
\textsf{U-I}  &  16   &  3   \\
\textsf{R-I}  & 8   &  15   \\
\textsf{I-A}  &  0   &  0   \\
\textsf{N-I}  & 1  & 1 \\ \hline

\end{tabular}}
\label{tab:categories-of-failures-rq4}
\end{table}

\noindent \textbf{Results.} The repairing results of LLM-CompDroid are shown in Table \ref{tab:repairing-result-rq4} with option \blackding{1} and option \blackding{2}. Where, option \blackding{2} has similar results to option \blackding{1} checked by our carefully manual analysis (the lower \textit{Correct} value of 90.9\% to LLM-CompDroid-GPT-4 \blackding{2} is due to that it has more unrepaired bugs, compared to other three LLM-CompDroid models, related to the attribute \code{android:gravity} which are discussed later in this section), thus, we mainly focus on option \blackding{1}'s results. Both GPT-3.5 and GPT-4-based LLM-CompDroid perform well. LLM-CompDroid-GPT-3.5 achieves 91.9\% \textit{Correct} value and 93.5\% \textit{Correct@k} value, and LLM-CompDroid-GPT-4 achieves 92.5\% 
 \textit{Correct} value and 96.1\% \textit{Correct@k} value. These results indicate that LLM-CompDroid is stable for repairing bugs since the gaps between \textit{Correct} values and \textit{Correct@k} values are both lower than 3.6\%. Moreover, the \textit{Overfitting} values of LLM-CompDroid-GPT-3.5 and LLM-CompDroid-GPT-4 are close to zero, with 1.6\% and 2.6\%, respectively. The \textit{Overfitting} is due to both LLMs themselves and ConfFix APIs~\cite{Huang:2023} in the cases of out-of-multi-round budget. ConfFix also has randomness, so the generated numerical attribute values may be different among different running, which leads to \textit{Overfitting}.

 We also count the failures by manual analysis, shown in Table \ref{tab:categories-of-failures-rq4}. All failures are in \textsf{U-I}, \textsf{R-I}, and \textsf{N-I}. Moreover, we also find that most of the failures are related to attribute \code{android:gravity}. LLM-CompDroid-GPT-3.5 and LLM-CompDroid-GPT-4 account for 67.9\% and 55.6\% for it, respectively. This result is interesting since LLM-CompDroid can provide valid candidate attributes (e.g., \code{android:top}), related to location information and used by app project authors for repairing bugs, to inside LLMs for repairing but both GPT-3.5 and GPT-4 mainly (60\%\footnote{All related cases are considered but one bug is not counted in because it has only one run rejecting candidate attributes, which is an outlier. For the following same calculation, we maintain this setting because they have the same trend.}) reject adopting them in these failure bug cases. Even if they are used, it is still not possible (1/11 probability) to generate correct location-related and numerical values. By observing option \blackding{2}'s result provided with fine-granularity location-related key field information, we find that inside LLMs still mainly (58\%) reject adopting them, and it is not possible (0/14 probability) to generate correct values by adopting them. Thus, we perform a case study on this problem in our discussion (Sec. \ref{sec:case-study}). For the rest failures, LLM-CompDroid-GPT-3.5 cannot repair 2 bugs and has another 2 bugs that each have 1 failure (caused by randomness). LLM-CompDroid-GPT-4 cannot repair 1 bug and has another 4 bugs that each have 1 or 2 failures. This result indicates that LLM-CompDroid can successfully repair bugs by prioritizing correct patches over failures. For the bugs that cannot be repaired by LLM-CompDroid-GPT-3.5, one is related to \code{android:gravity}, and another one is related to \code{android:background}. For the latter LLM-CompDroid-GPT-3.5 changes it to \code{android:src}, which causes another configuration compatibility bug related to issue-inducing attribute \code{android:src}. Unfortunately, LLM-CompDroid cannot provide new bug-related information to GPT-3.5 to avoid it, causing failures. However, LLM-CompDroid-GPT-4 can successfully generate repairs for this bug by replacing \code{android:background} with \code{app:srcCompat} in most cases where \code{app:srcCompat} is also an issue-fixing attribute for \code{android:src}. For LLM-CompDroid-GPT-4, it fails to repair \code{android:layout\_height}-related bug (LLM-CompDroid-GPT-4 changes values to numerical values which cannot repair bugs) but LLM-CompDroid-GPT-3.5 and both pure LLMs can repair it. These results indicate that different LLMs used can have different impacts on repairing results in LLM-CompDroid framework. 

\begin{tcolorbox}[boxrule=1pt,boxsep=1pt,left=2pt,right=2pt,top=2pt,bottom=2pt]
\textbf{Answer to RQ4:}
LLM-CompDroid achieves high performance in repairing configuration compatibility bugs under LLMs of GPT-3.5 and GPT-4. Specifically, LLM-CompDroid-GPT-3.5 achieves 91.9\% \textit{Correct} value and 93.5\% \textit{Correct@k} value, and LLM-CompDroid-GPT-4 achieves 92.5\% \textit{Correct} value and 96.1\% \textit{Correct@k} value.
\end{tcolorbox}

\subsection{RQ5: How does LLM-CompDroid compare to traditional tools in repairing bugs?}\label{subsec:req5}

\noindent\textbf{Motivation.} We intend to compare LLM-CompDroid's performance with other baseline methods.

\noindent \textbf{Methodology.} To answer this RQ, we run the latest Lint version (lint: version 8.1.0)~\cite{Lint} on the 77 configuration compatibility bugs. For each bug, we run Lint once due to its a rule-based static analyzer with no randomness. In addition, we continue to use the results of pure GPT-3.5 and GPT-4 from Sec. \ref{subsec:rq3}. Furthermore, for LLM-after-ConfDroid, we first run ConfFix to repair bugs. If ConfFix fails to repair bugs, then we trivially query LLMs for repairing these unrepaired bugs from ConfFix by leveraging the same prompts used in Sec. \ref{subsec:rq3}. The evaluation metrics used include: \textit{Correct}, \textit{Overfitting}, and \textit{Correct@k}.

\noindent\textbf{Results.} The experimental results are shown in Table. \ref{tab:repairing-result-rq4}. Both LLM-CompDroid-GPT-3.5 and LLM-CompDroid-GPT-4 outperform pure GPT-3.5 and GPT-4 with 38.1\% to 77.2\% advantage in \textit{Correct} and 19.5\% to 64.9\% advantage in \textit{Correct@k}. Moreover, we count the mean values of \textit{Correct} for those bugs repaired dependent on LLMs in LLM-CompDroid. The same bugs' are also counted in both pure GPT-3.5 and GPT-4. For LLM-CompDroid-GPT-3.5 and LLM-CompDroid-GPT-4, they achieve 0.63 and 0.66 mean values, respectively. As for pure GPT-3.5 and GPT-4, they achieve 0.29 and 0.4 mean values, respectively. This result shows that LLM-CompDroid can increase the stability of repairing by LLMs by 0.34 and 0.26, respectively. In addition, LLM-CompDroid-GPT-4 also repairs 2 more hard-to-repair bugs than pure GPT-4 due to the candidate attributes provided. GPT-4 in LLM-CompDroid is able to refer to these candidate attributes for repairing instead of searching for patches in the whole attribute search space. Moreover, LLM-CompDroid is also better than LLM-after-ConFix. Both LLM-after-ConfFix-GPT-3.5 and LLM-after-ConfFix-GPT-4 have \textit{Correct} values lower than 90\%, with 87.5\% and 89.4\%, respectively. This result shows again that LLM-CompDroid can help LLMs generate more correct repairs. However, LLM-after-ConfFix-GPT-3.5 and LLM-after-ConfFix-GPT-4 have one more repairing to one bug (repaired by ConfFix) than LLM-CompDroid-GPT-3.5, resulting in higher \textit{Correct@k} values with 94.8\%. This bug is related to the attribute \code{android:gravity} which is discussed in Sec. \ref{sec:case-study}.

For Lint in the latest version, it successfully generates 32 patches, achieving 41.6\% \textit{Correct} without any \textit{Overfitting}. For the bugs repaired by Lint, LLM-CompDroid can also repair them, attributed to LLM-CompDroid's ConfFix APIs. However, Lint can only repair one kind of bug related to \code{android:drawableTint}, an easy-to-repair bug. This is because Lint is a rule-based tool that does not contain other bugs' repairing information and other bugs cannot be repaired. In conclusion, LLM-CompDroid outperforms Lint.

As for ConfFix, it achieves 82.1\% \textit{Correct} value and 83.1\% \textit{Correct@k} value with a low \textit{Overfitting} in 1.0\%. ConfFix outperforms Lint and two pure LLMs but is worse than both LLM-CompDroid-GPT-3.5 and LLM-CompDroid-GPT-4 by 9.8\% and 10.4\% in \textit{Correct} values and 10.4\% and 13.0\% in \textit{Correct@k} values. ConfFix can repair easy-to-repair bugs, however, when it encounters bug repairs that do not adhere to established rules or the defined fitness function (e.g., issue-fixing attributes or elements contribute nothing to key fields), ConfFix fails to repair them. Nevertheless, LLM-CompDroid can leverage the flexibility of LLMs and three roles to converge and complete the repair of hard-to-repair bugs. For example, when repairing the bug related to <\code{ImageView}> element and \code{android:foreground} attribute (see Fig. \ref{fig:example-compatibility-bug-intro}), LLM-CompDroid-GPT-4's Repairer generates a patch that replaces \code{android:foreground} with \code{app:srcCompat} which is not a correct issue-fixing attribute and Checker finds the problem, telling Repairer the repair is incorrect with explanation. So, Repairer generates another patch by introducing <\code{FrameLayout}> which is a correct one. Subsequently, both Checker and Optimizer pass the repair proposal, and LLM-CompDroid-GPT-4 generates the successfully repaired app. There is another example related to \textit{seekbar} bug though LLM-CompDroid-GPT-4 finally fails to generate a correct patch. LLM-CompDroid-GPT-4's Repairer generates a patch by replacing \code{android:layout\_height} with \code{android:minHeight}. Checker finds it is not a correct patch and provides two options to Repairer with a correct patch \code{android:layout\_height="wrap\_content"} and an overfitting patch \code{android:layout\_height="100dp"}. Unfortunately, Repairer chooses the latter one and both Checker and Optimizer pass the repairing proposal. Though this is a failure, it still demonstrates the potential to repair hard-to-repair bugs by LLM-CompDroid. This is also a reason why LLM-CompDroid can perform more stably. However, ConfFix cannot find any attributes contributing to the fitness function for repairing these bugs, resulting in failures. 

Moreover, we also count the iterative interaction among the three roles of LLMs in practice. For the hard-to-repair bugs which have 15 ones (75 times for repairing in the experiment), there are 24 and 20 times the iterative interaction loops are executed more than once (the first loop repairs are incorrect) to repair a bug, for LLM-CompDroid-GPT-3.5 and LLM-CompDroid-GPT-4, respectively. Where there are 11 (45.8\%) and 4 (20\%) times for LLM-CompDroid-GPT-3.5 and LLM-CompDroid-GPT-4 respectively to have successful repairs. Additionally, there are 1.41 and 1.56 repetitions in the iterative interaction loop process for LLM-CompDroid-GPT-3.5 and LLM-CompDroid-GPT-4 respectively on average.

Therefore, we can conclude that the LLM-CompDroid framework can combine traditional tools and LLMs to effectively and stably repair configuration compatibility bugs, especially for hard-to-repair bugs.

\begin{tcolorbox}[boxrule=1pt,boxsep=1pt,left=2pt,right=2pt,top=2pt,bottom=2pt]
\textbf{Answer to RQ5:}
Both LLM-CompDroid-GPT-3.5 and LLM-CompDroid-GPT-4 outperform traditional tools, pure LLMs, and LLM-after-ConfFix in repairing configuration compatibility bugs. In addition, LLM-CompDroid can combine traditional tools and LLMs to effectively and stably repair configuration compatibility bugs, especially for hard-to-repair bugs (e.g., bugs unrelated to fitness function).
\end{tcolorbox}

\section{Discussion}\label{sec:Discussion}

\subsection{Case Study: android:gravity Problem}\label{sec:case-study}



\begin{figure}[t]
    \centering
    \subfigure{
    \includegraphics[width=0.32\textwidth]{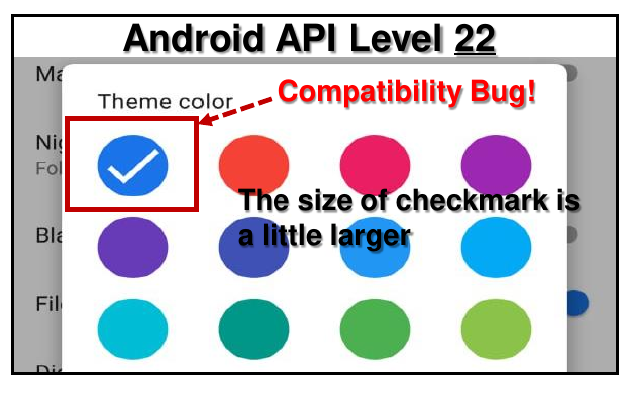}}
    \subfigure{
    \includegraphics[width=0.32\textwidth]{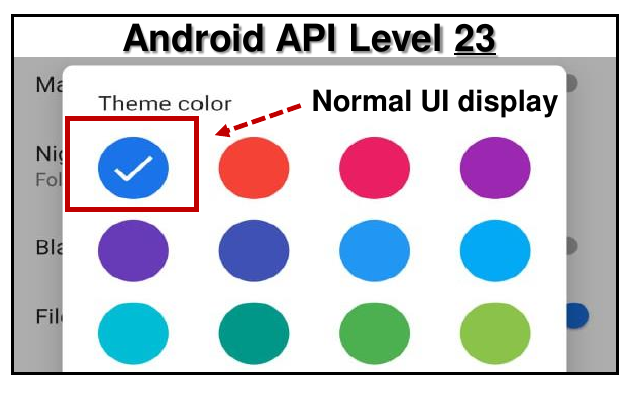}}
    \subfigure{
    \includegraphics[width=0.32\textwidth]{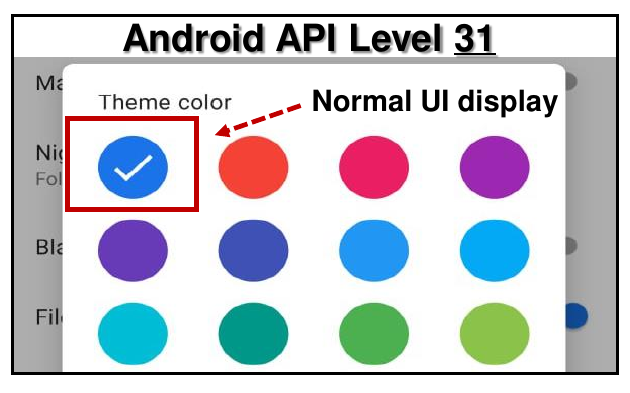}}
    \caption{Example with low impact level across Android API levels 22, 23, and 31.}
    \label{fig:example-compatibility-bug-diss}
\end{figure}

From the results in Sec. \ref{subsec:req4} for \code{android:gravity}-related configuration compatibility bugs, we find that in most of the cases, LLMs reject adopting the candidate attributes to \code{android:gravity} and even the candidate attributes are adopted, it is still not possible for LLMs to generate correct location-related and numerical values for repairing. Here, we discuss this problem. For each related bug, we perform 5 more repairing for them and the result shows that there are still 50\% cases that reject adopting candidate attributes. Moreover, we also observe related bugs in the responses of pure GPT-3.5 and GPT-4 and find that 74\% of them reject introducing other location-related attributes. Regardless of whether additional candidate repair information and roles (sometimes, Checker even may prevent Repairer from using candidate attributes) are provided, LLMs generally reject introducing other location-related attributes. Based on this result, we can conclude that prompts may not be the reason for LLMs to reject the use of location-related attributes. Therefore, we check the runtime behavior (caused by the configuration compatibility bugs across conflicting Android API levels and the maximum Android API level) of these apps and find that the impact of 1 bug is at high level (cause crashing), the impact of 2 bugs is at medium level (slightly affect usability. The entered 10-digit number is obscured), and the impact of another 3 bugs is at low level (does not affect the use at all. e.g., see Fig. \ref{fig:example-compatibility-bug-diss}) which may be seen as approximately false positives to humans. For the first bug, all repairs introduce location-related attributes except one. However, for other bugs, they are in the \code{android:gravity} problem. Through this result, one possible reason for this problem can be concluded that for the latter 5 bugs with their context XML elements, their impact on user usage may be minimal, so the relevant XML code on the internet may be predominantly unrepaired cases. As for the first 1 bug with its context XML elements, it causes crashing and has a great impact on users, resulting in having predominantly repaired cases, and thus, LLMs can repair the bug by using location-related attributes to maintain consistency of visual effects across different Android API levels. Therefore, future-designed configuration compatibility bug detection tools should also consider the context information of issue-inducing XML elements and issue-inducing attributes in relevant scenarios to detect bugs, in order to avoid detecting approximately false positives like the 3 bugs at low level above. Moreover, future-designed LLM or traditional-based repairing tools also should consider how to effectively resolve bugs at medium level like the 2 ones above by introducing location-related attributes with appropriate values.

\subsection{Lessons Learned}
In this section, we discuss the lessons learned from this study.

\noindent \textbf{Configuration Compatibility Bug Detection.} LLMs are not effective in detecting configuration compatibility bugs in Android apps. They may introduce many false positives and negatives in detection tasks. The possible reason is that there is less bug-relevant data in the training dataset (see Sec. \ref{sec:d-ll}).

\noindent \textbf{Configuration Compatibility Bug Repairing.} Though, in general, LLMs perform weakly and unstably under repairing tasks, they show capabilities in repairing hard-to-repair bugs. Thus, this insight gives a chance to integrate LLMs with traditional repairing tools (e.g., ConfFix~\cite{Huang:2023}) to leverage the advantages from both sides. Experimental evaluation shows that LLM-CompDroid improves the repairing performance significantly. We also believe that this insight and conclusion can potentially contribute to the development of LLMs in other related fields.

\subsection{Threats to Validity}

\noindent\textbf{External Validity.} To reduce the potential bias in building a dataset of configuration compatibility bugs, we reuse the previously studied dataset~\cite{Huang:2023}. This dataset contains 77 configuration compatibility bugs in 13 Android apps, covering diverse issue-inducing attributes. Moreover, the 13 Android apps are also diverse in multiple app categories
and popular with thousands to millions of downloads in Google Play. Each of them is with rich maintenance and most of them have over 1,000 stars on GitHub. However, the size of the dataset is relatively small, thus, we do not attempt to generalize our results and conclusions to all configuration compatibility bugs in Android apps.

\noindent\textbf{Internal Validity.} The threat to internal validity is the randomness of LLMs. To reduce this risk, we set a temperature of 0 to LLMs in detection tasks, and run 5 times with a temperature of 0.7 for each configuration compatibility bug in repairing tasks.

\section{Related Work}\label{sec:RelatedWork}

\subsection{Detecting Compatibility Bugs in Apps}
Detecting compatibility bugs in Android apps has been widely studied in the existing research works \cite{Huang:2018,Huang:2021,He:2018,Fazzini:2017,Ki:2019,Li:2018,Li:2018cid,Liu:2022,Wei:2016,Wei:2018,Wei:2019}. Specifically, DiffDroid \cite{Fazzini:2017} and Mimic \cite{Ki:2019} adopt dynamic-based approaches to detect compatibility bugs by comparing the app UI differences among different Android devices. DiffDroid \cite{Fazzini:2017} combines input generation and differential testing to compare the behavior of an app on different platforms and identify possible inconsistencies. Mimic \cite{Ki:2019} is designed
specifically for comparing the UI behavior of an app across
different devices and Android versions. Contrast to dynamic-based approaches, some works \cite{Huang:2018, Huang:2021, He:2018, Li:2018, Li:2018cid, Liu:2022, Wei:2016, Wei:2018, Wei:2019} leverage static-based approaches (i.e., pre-defined rules) to detect compatibility bugs in apps. For example, ConfDroid \cite{Huang:2021} is driven by the rules extracted from the Android framework code changes to detect compatibility issues. Cid \cite{Li:2018} can systematically model the lifecycle of the Android APIs and analyze app bytecode to flag usages that can lead to potential compatibility issues. Pivot \cite{Wei:2019} can automatically learn API-device correlations of compatibility issues from existing Android apps. It extracts and prioritizes API-device correlations from a given corpus of Android apps.

\subsection{Repairing Compatibility Bugs in Apps}
To repair the compatibility bugs in apps, researchers propose various automatic solutions for repairing compatibility bugs~\cite{Huang:2023, Zhao:2022, Lamothe:2020, Haryono:2020, Haryono:2022androevolve, Fazzini:2019}. For example, Huang et al.'s work \cite{Huang:2023} extracts knowledge learned from the Android framework to facilitate configuration compatibility issue repair. Zhao et al.'s tool \cite{Zhao:2022} RepairDroid offers a generic app path description language for users to create fix templates for compatibility issues. The templates will then be used by RepairDroid to fix the compatibility bugs at the bytecode level. Haryono et al.'s tool  \cite{Haryono:2022androevolve} AndroEvolve analyzes data flows in a file scope to infer more accurate transformations. Lamothe et al. \cite{Lamothe:2020} 's work A3 updates the usage of incompatible APIs based on the patterns learned from the code examples in other apps. Android's official toolkit Lint \cite{Lint} is a static analyzer to check resource usages. It contains predefined rules for repairing configuration compatibility issues.

\subsection{Large Language Model Studies}
There are several research working on LLM-based detection and repairing tasks~\cite{xia:2023, fan2023automated, pearce:2023, sun2023gpt, Dipongkor:2023, liu2023not, lee2022light}. Fan et al.~\cite{fan2023automated} conduct a comprehensive investigation into the effectiveness of automated program repair (APR) techniques, including Codex~\cite{chen2021evaluating}, in rectifying erroneous solutions generated by Codex for LeetCode problems. Xia et al.~\cite{xia:2023} conduct an extensive examination of the direct utilization of LLMs, specifically 9 state-of-the-art LLMs, for APR. Pearce et al.~\cite{pearce:2023} investigate the application of LLMs, such as Codex, in the context of zero-shot learning for vulnerability repairing. Sun et al.~\cite{sun2023gpt} leverage static analysis and LLMs to detect smart contract logic vulnerabilities. Dipongkor et al.~\cite{Dipongkor:2023} conduct a comprehensive study on bug triaging by leveraging Transformer-based techniques. Similar work also includes~\cite{lee2022light}. There is also much work studying LLM-based fuzzing to detect bugs in software~\cite{tang:2023, lemieux2023codamosa, deng2023large, liu2023fill}. Different from these studies, our research focuses on detecting and repairing configuration compatibility bugs in Android apps with LLMs.
\section{Conclusion}\label{sec:Conclusion}


In this study, we conduct a thorough examination of large language model (LLM)-based approaches for detecting and repairing configuration compatibility bugs. Our findings reveal that LLMs are ineffective at detecting these bugs and generally exhibit weaknesses in bug repairing, although they demonstrate promise in addressing complex issues. Leveraging these insights, we introduce the LLM-CompDroid framework, combining LLMs and traditional tools to significantly enhance bug resolution. Specifically, LLM-CompDroid-GPT-3.5 and LLM-CompDroid-GPT-4 outperform ConfFix by at least 9.8\% and 10.4\% in both \textit{Correct} and \textit{Correct@k} metrics. Our study's implications extend beyond this domain, potentially contributing to the broader development of LLMs in related fields within the research and innovation landscape.

\normalem
\balance

\bibliographystyle{ACM-Reference-Format}
\bibliography{refs/refs}

\end{document}